\documentclass[ prd,  nofootinbib]{revtex4}
\usepackage{amssymb,amsmath,graphicx,color, microtype}
\usepackage{epsf}
\usepackage{epstopdf}
\usepackage {amssymb}
\newcommand{\nc}{\newcommand}
\nc{\ba}{\begin{eqnarray}}
\nc{\ea}{\end{eqnarray}}
\newcommand\be{\begin{equation}}
\newcommand\ee{\end{equation}}

\usepackage{color}

\nc{\e}{{\bf{e}}}
\nc{\kk}{{\bf{k}}}
\nc{\pp}{{\bf{p}}}

\nc{\bfk}{{\bf{k}}}
\nc{\bfx}{{\bf{x}}}
\nc{\bfp}{{\bf{p}}}

\nc{\eH}{{\epsilon_H}}
\nc{\calP}{{\cal P}}
\nc{\im}{{ \mathrm{Im} } }

\begin{document}
\title{Clustering Fossil from Primordial Gravitational Waves in Anisotropic Inflation}

\author{Razieh Emami$^{1}$}
\email{emami-AT-ipm.ir}

\author{Hassan Firouzjahi$^2$}
\email{firouz-AT-ipm.ir}

\affiliation{$^1$ School of Physics, Institute for Research in
	Fundamental Sciences (IPM)
	P.~O.~Box 19395-5531,
	Tehran, Iran \\
	$^2$School of Astronomy, Institute for Research in
	Fundamental Sciences (IPM)	P.~O.~Box 19395-5531,
	Tehran, Iran}

\begin{abstract}

Inflationary models can correlate small-scale density perturbations with the long-wavelength gravitational waves (GW) in the form of the Tensor-Scalar-Scalar (TSS) bispectrum. This correlation affects the mass-distribution in the Universe and leads to the off-diagonal correlations of the density field modes in the form of the quadrupole anisotropy. Interestingly, this effect survives even after the tensor mode decays when it re-enters the horizon, known as the  fossil effect. As a result, 
the off-diagonal correlation function between different Fourier modes of the density fluctuations can be thought as a way to probe the large-scale GW and the 
mechanism of inflation behind the fossil effect. 
Models of single field slow roll inflation generically predict a very small quadrupole anisotropy in TSS while in models of multiple fields inflation this effect can be observable.  Therefore this large scale quadrupole anisotropy can be thought as a spectroscopy for different inflationary models.
In addition, in  models of anisotropic inflation there exists  quadrupole anisotropy in curvature perturbation power spectrum.   
Here we consider TSS in models of  anisotropic inflation and show that the shape of quadrupole anisotropy is different than in single field models. In addition 
in these models the quadrupole anisotropy is projected into the preferred direction and its amplitude is proportional to $g_* N_e$ where $N_e$ is the number of e-folds and $g_*$  is the amplitude of quadrupole anisotropy in curvature perturbation power spectrum. We use this correlation function to estimate the large scale GW as well as the preferred direction and discuss the detectability of the signal in the galaxy surveys like Euclid and 21 cm surveys. 

\end{abstract}
\maketitle

\section{Introduction}

The recent cosmological observations such as Planck 2015  are in very good agreement with the inflationary cosmology as  the source of the density perturbations in the CMB and the origins of  structures in the Universe \cite{Ade:2015lrj}. Although CMB observations are consistent with the single field slow-roll (SFSR) inflation models but there can be other types of the fields including additional scalar, vector and tensor modes which could play a role during inflation in a distinguishable way \cite{Jeong:2012df}. Therefore,  it is  worth to search for their observational imprints. Among the above additional fields, tensor modes are very interesting because they are one of the inevitable predictions of  general relativity. There have been some experiments searching  for the imprints of the gravitational waves in the CMB \cite{Ade:2014xna, Ade:2015tva, Keisler:2015jua} however there was no detection of primordial tensor mode. 
From the Planck results an upper bound $r\lesssim 0.1$ for the ratio of the amplitude of tensor perturbation to scalar  perturbation in CMB power spectrum is obtained.  
While the CMB experiments are near to their completion,  it is worth to search for primordial GW in  large-scale structure surveys like Euclid and 21-cm surveys. There were also some theoretical works looking for the imprints of the GW in the CMB and large-scale surveys such as lensing by the tensor modes of the galaxy distribution, 21-cm fluctuations etc.  \cite{Dodelson:2003bv, Schmidt:2012nw, Smith:2008pf, Pagano:2007st, Smith:2006nka, Pritchard:2004qp, Kesden:2004qx, Kesden:2002ku, Kamionkowski:2000gb, Caldwell:1998aa, Kamionkowski:1997av, Dai:2012bc, Chisari:2014xia, Pen:2003yv, Masui:2010cz, Book:2011dz, Akhshik:2014gja}. 

The key point in the galaxy searches is the two-point correlation function or power-spectrum. Under the null hypothesis, this two-point function is statistically homogeneous and isotropic. In addition, there is no correlation between different Fourier modes. Coupling the inflaton field with the tensor mode leads to off-diagonal correlation between different Fourier modes and thus can be thought as a way to probe the primordial GW. However, since the final goal of the galaxy surveys is to look at the power-spectrum of the density perturbation, it is convenient to assume that there is a hierarchy between the scale of the density perturbations, supposed to have small wavelength, and that of the tensor mode, assuming to have a long wavelength. Therefore one should consider the three-point function of the Tensor-Scalar-Scalar (TSS) at the squeezed limit. It turns out that this three-point function looks like a quadrupole anisotropy in power spectrum. 

The next step is considering the non-linear evolution of this bispectrum during the radiation and matter dominated 
eras and finally, in order to read the observed mass distribution, it is also required to consider the projection effects \cite{Dai:2013kra, Schmidt:2013gwa, Pajer:2013ana}. Intuitively, this effect comes from the fact that in a redshift survey one can find the position of the galaxies by using their apparent position and redshift in the sky. However, in the presence of the tensor mode, the apparent position will not be the same as the original position of the object. As a result there are three different potentially important contributions that must be taken into account in order to evaluate the observed power spectrum of the mass distribution in the Universe. However, among the above effects, only the first one depends on the specific inflationary model. So one can neglect the non-linear and projection effects and consider only the primordial bispectrum. 

It has been shown that for  SFSR inflation models, all of the above effects are at the same order and there would be a cancellation between them. This seems to be the case for all models which satisfy the Maldacena's consistency condition \cite{Maldacena:2002vr}. As a result, for these kinds of models, the signal would be small and beyond the accessibility of the current galaxy surveys. So it is worth to go beyond the SFSR inflation and see how the violation of the consistency relation shows up in the observed power spectrum. This has been done very recently in  \cite{Dimastrogiovanni:2014ina, Akhshik:2014bla, Dimastrogiovanni:2015pla}. As they have shown there are no cancellation between different components  in these models. The authors have claimed that the quadrupole amplitude is observable by the current galaxy surveys. However, its shape is exactly the same as in SFSR inflation models. 

It is well-known that the GW has a nontrivial evolution, an oscillatory and decaying profile, when it re-enters the horizon. However, it can be shown that the above phenomenon is a cumulative effect and does survive even after the long wavelength tensor mode decays inside the horizon \cite{Dai:2013kra}. As a result this effect 
is called the \textit{fossil effect}.

In this paper, we  generalize these studies to models of anisotropic inflation in which there is a preferred direction 
during inflation yielding to statistical anisotropies in the form of quadrupole anisotropy in curvature perturbation power spectrum. In addition, the bispectrum has a different shape and the TSS would be aligned along that direction in contrast to SFSR models.   In models of  anisotropic inflation we assume that at the background level there is a vector field which sources the background anisotropy.  As we shall see in the following this anisotropy affects the TSS and we will obtain 
\ba
\label{final result}
\langle h^{(\lambda)}_{\mathbf{k_1}} \zeta_{\mathbf{k_2}} \zeta_{\mathbf{k_3}} \rangle  = \frac{3g_{*}}{2}  N_e \left(\mathbf{\hat{x}} \cdot e^{(\lambda)}(\mathbf{k_1})\cdot \mathbf{\hat{x}} - \mathbf{\widehat{k}_3}\cdot \mathbf{\hat{x}} \, \mathbf{\hat{x}}\cdot e^{(\lambda)}(\mathbf{k_1})\cdot \mathbf{\widehat{k}_3} \right) P_{\zeta}(\mathbf{k_3}) P_{ h^{(\lambda)}}(\mathbf{k_1}) + (\mathbf{k_3} \leftrightarrow \mathbf{k_2}) 
\ea
where $\mathbf{\hat{x}}$ refers to the preferred direction and $\lambda$ is for the tensor polarizations, $+,\times$. Note that  a  $\cdot$ denotes the inner product between a tensor mode, associated with the polarization of the gravitational wave $ e^{(\lambda)}$  which has two indices, and two vectors. For example
$ \mathbf{\hat{x}} \cdot e^{(\lambda)}\cdot \mathbf{\hat{k}} =  \mathbf{\hat{x}_i} \, e^{(\lambda)}_{ij}  \, \mathbf{\hat{k}_j} $.  We should emphasis here that the above shape is the leading shape for the TSS.
There are also some other terms which contribute in TSS. However, as we will show, their contribution would be sub-leading. 

As we will discuss, in models of anisotropic inflation there will be quadrupole anisotropy in two point function
induced from  the preferred direction so this TSS would be added on the top of that anisotropy.  
We  try to use this anisotropy as a way to probe the tensor mode as well as to find the preferred direction. This can be done by writing the optimal estimator for GW. As we will show, there is a window in our parameter space which can yield signals detectable by Euclid and/or 21 cm surveys. 

The rest of this paper is organized as following. In Section \ref{model} we present the 
setup of anisotropic inflation in some details. In Section \ref{interactions} we  calculate the TSS correlation function. Next, in Section \ref{observation}, we consider the observational consequences of these analysis 
for the galaxy surveys. We end with  Conclusion and leave some technical details into appendices.

\section{Anisotropic Inflation}
\label{model}

In this section we briefly review the setup of anisotropic inflation. For related works on anisotropic inflation see  \cite{Watanabe:2009ct, Watanabe:2010fh, Soda1, Emami1, Emami:2013bk, Abolhasani:2013zya, Abolhasani:2013bpa,  Chen:2014eua, Bartolo:2012sd, Shiraishi:2013vja, Shiraishi:2013oqa,  various, Kim:2013gka}, for a review  see \cite{Soda:2012zm, Maleknejad:2012fw}. 

\subsection{The background }
\label{background}

Our anisotropic inflation model contains a $U(1)$ gauge field $A_\mu$ with non-zero amplitude at the background level. However, as it is well-known, the Maxwell theory suffers from the conformal invariance in an expanding backgrounds. As a result, the background gauge field $A_\mu(t)$ is diluted exponentially during  inflation. Moreover,  the quantum excitations of the gauge field $\delta A_\mu (t, \bfx)$ during inflation 
will not be scale-invariant. Therefore,  it is essential to break the conformal invariance.  The easiest method to break the conformal invariance is to couple the gauge field to inflaton field to develop a time-dependent gauge kinetic coupling.   With this discussion in  mind the action of our system is 
\ba
 \label{action} S= \int
d^4 x  \sqrt{-g} \left [ \frac{M_P^2}{2} R - \frac{1}{2} \partial_\mu \phi
\, \partial^\mu \phi -   \frac{f^{2}(\phi)}{4} F_{\mu \nu} F^{\mu
\nu}  - V(\phi) \right] \, ,
\ea
where $M_P$ is the reduced Planck mass while $f(\phi)$ represents the time-dependent gauge kinetic coupling which couples the inflaton field $\phi$ to the gauge field. 
As usual, $F_{\mu \nu}$ is the gauge field strength which is given by
\ba F_{\mu \nu} = \nabla_\mu A_\nu
- \nabla_\nu A_\mu  = \partial_\mu A_\nu - \partial_\nu A_\mu \, .
\ea
The model of anisotropic inflation based on the above action was studied at the background level in \cite{Watanabe:2009ct} and its  analysis for curvature perturbations were performed in \cite{Watanabe:2010fh, Emami:2013bk, Abolhasani:2013zya, Abolhasani:2013bpa, Chen:2014eua,   Bartolo:2012sd, Shiraishi:2013vja, Shiraishi:2013oqa}.

Note that in  Maxwell theory $f(\phi)=1$. However, as mentioned above, 
in order to break the conformal invariance, we need a time-dependent gauge kinetic coupling. As it has been shown in  \cite{Watanabe:2009ct, Watanabe:2010fh, Soda1, Emami1, Emami:2013bk,  Abolhasani:2013zya, Abolhasani:2013bpa,  Chen:2014eua,  Bartolo:2012sd, Shiraishi:2013vja, Shiraishi:2013oqa}, one requires $f \propto a^{-2}$ in order to obtain a scale invariant power spectrum for the gauge field quantum fluctuations. With this choice of the gauge kinetic coupling one obtains a constant background electric field energy density during the inflation. 

We choose the convention that  the background gauge field is turned on along the $x$-direction  so  $A_\mu = (0, A_x(t), 0,0)$. With this choice for the gauge field, the background space-time 
still poses the symmetry in $y-z$ plane. The background is in the form of Bianchi I universe
with the metric given by 
\ba
\label{bian0}
ds^2 &=& -dt^2 + a(t)^2 dx^2 + b(t)^2 \left(dy^2 + dz^2\right) \nonumber\\
&=&- dt^2 + e^{2\alpha(t)}\left( e^{-4\sigma(t)}d x^2
+e^{2\sigma(t)}(d y^2 +d z^2) \right) \, .
\ea
Here $\dot \alpha$ measures the average of the Hubble expansion rate while $\dot \sigma(t)$ measures the level of anisotropic expansion.  However, the observational constraints on the level of anisotropy in curvature perturbations is not more than few percent  \cite{Kim:2013gka} so 
$\dot \sigma /\dot \alpha \ll 1$. As a result, one can treat the analysis perturbatively in this model.
 
The background fields equations have been studied in  \cite{Watanabe:2009ct}. It is shown that 
with the general gauge kinetic coupling 
\ba
\label{f-form0}
f(\phi) = \exp {\left( \frac{c\phi^2}{2 M_P^2}  \right)} =  \left(\frac{a}{ a_{\mathrm{end}}}\right)^{-2c}\, ,
\ea
the system reaches an attractor regime for $c >1$ in which the ratio of the gauge field energy density $\rho_A$ (i.e.  the electric field energy density) becomes a small but constant fraction of the total energy density $\rho$. It is convenient to define   the fraction of the gauge field energy density  to total energy density via
\ba
\label{R-der}
R \equiv  \frac{\rho_A}{\rho} = \frac{\dot A_x^2 f(\phi)^2 e^{-2 \alpha}}{2V} \, .
\ea
As demonstrated in  \cite{Watanabe:2009ct}, during the attractor phase one obtains
\ba
\label{R-attractor}
R = \frac{I \epsilon_H}{2} 
\ea
in which $I \equiv \frac{c-1}{c}$ is a measure of anisotropy in the system and $\epsilon_H \equiv -\dot H/H^2$ is the usual slow-roll parameter.  The bound on $R$
is determined by the quadrupole anisotropy in curvature power spectrum as we discuss below. 

\subsection{Perturbations}
\label{perturbation}

In the following, we review perturbations in anisotropic inflation setup.  
We  divide the analysis into three parts:  the quadratic action, the second order interaction terms as well as the third order interactions.

In general, the perturbations in anisotropic setup originate both from the metric sector as well as  from the matter sector. Specifically, the metric sector has the  non-dynamical degrees of freedom $\delta g_{0 \mu}$ which should be integrated out in order to calculate the dynamical action. After integrating out these non-dynamical metric degrees of freedom one obtains many new terms in the dynamical action. However, as it has been verified  in \cite{Emami:2013bk},  the perturbations from the 
metric are  either slow-roll suppressed or would cancel with each other. Therefore,  to leading order in slow-roll parameters and the anisotropic parameter $R$,  the leading terms in the interactions come entirely from the matter sector perturbations. In other word,  we do not have to consider the perturbations from the non-dynamical degrees of freedom in the metric which simplifies the analysis considerably. Technically speaking this corresponds to the decoupling limit in which one can neglect the gravitational back-reactions with the expense of  inducing errors at the order of slow-roll parameters.

To simplify the analysis further, we choose the flat gauge in which the curvature perturbations is given by the inflaton perturbations $\zeta = - \frac{H}{\dot \phi} \delta \phi$. As a result, the metric takes the following simple form,
\ba
\label{dynamical metric}
ds^2 = a(\eta)^2 \left(-d\eta^2 + \left[\delta_{ij}+ h_{ij} \right]dx^i dx^j \right ) \, ,
\ea
where $\eta$ represents the conformal time $d\eta = dt/a(t)$. 
As  mentioned above, since the leading corrections come from the matter fluctuations, one can safely set $a=b$; the errors induced with this assumption are higher orders in $R$ and $\epsilon_H$.  The perturbations $h_{ij}$ above represent the tensor modes which are transverse and traceless: $\partial_i h_{ij} = 0$ and $ h_{ii}=0$  where the repeated indices are summed over.  We denote the two different polarizations of the metric by $h_\times$ and  $h_+$. Finally, for the gauge field perturbations, we choose the Coulomb-radiation gauge where $\delta A_{0}=0$ and $\partial_i \delta A_i =0$. 

We  quantize  the curvature perturbation, the gauge field perturbations and the tensor perturbations 
as usual. Let us start with the curvature perturbation  
\ba
\label{zeta-phi}
\zeta = - \frac{H}{\dot \phi} \delta \phi  = \frac{\delta \phi}{M_P \sqrt{2 \epsilon_H}}.
\ea
Going to the Fourier space, we can expand the curvature perturbation in terms of the annihilation and the creation operators $a(\bfk)$ and $a^{\dagger}(\bfk)$ as follows
\ba
 \zeta (\bfx , \eta) =  \int \frac{d^3k}{(2\pi)^{3/2}} e^{i \mathbf{k}.\mathbf{x}}
\widehat\zeta(\mathbf{k},\eta) \quad , \quad
\widehat\zeta(\mathbf{k},\eta) \equiv \zeta(k, \eta) a(\bfk) +   \zeta^*(k, \eta)  a^{\dagger}(- \bfk) \, ,
\ea
where, as usual, the creation and the annihilation operators satisfy the commutation relation
$ [ a(\bfk),    a^\dagger  (\bfk')] = \delta^{(3)} (\bfk-\bfk' )$.

In the limit that we neglect the gravitational back-reactions,  the wave function of the curvature perturbation corresponds to the profile of a  massless scalar field in a dS background,
\ba
\label{zeta-wave}
\zeta_k(\eta) = \frac{i H \eta }{M_{P}\sqrt{2\epsilon_{H}k} }  \left( 1 - \frac{i}{k\eta} \right) e^{-ik\eta} \, .
\ea
Here we have assumed the Bunch-Davies vacuum. It is also possible to relax this assumption, as it has been considered in \cite{Emami:2014tpa, Chen:2014vja}. However, in the following, we avoid this complication and simply assume that all fields start with the Bunch-Davies vacuum deep inside the horizon.

The curvature perturbation power spectrum  is given by,
\ba
\langle   \widehat{\zeta}(\mathbf{k_1})  \widehat{\zeta} (\mathbf{k_2}) \rangle
= (2 \pi)^3 \delta^{(3)} (\bfk_1 + \bfk_2) P_\zeta (k_1) \quad , \quad
\calP_\zeta \equiv   \frac{ k_1^3  }{ 2 \pi^2}  P_\zeta(k_1)
\ea
The isotropic power spectrum,  in the absence  of anisotropy, is
\ba
\label{calP0}
\calP_\zeta^{(0)} = \frac{H^2}{8 \pi^2 \epsilon_H M_P^2}\, .
\ea
We calculate the anisotropic power spectrum induced from the gauge field later on. 

To study the gauge field and tensor perturbations we have to specify the wave number of the modes as well as the polarizations of the vector and  the tensor fields. Although at the two point level it is enough (as usually assumed)  to take the wave number vector in the $x-y$ plane but this  is not the most general choice at the third order level.  
The most general wavenumber is given by  
\ba
\label{k-general}
\mathbf{k} = k\, \Big(\cos{\theta} \sin{\varphi}\, , \sin{\theta}\sin{\varphi}\, , \cos{\varphi} \Big), 
\ea
where $\phi$ is the angle of the wavenumber with respect to the $z$ direction and $\theta$ represents its angle with respect to the $x$ direction (note that the $x$-direction is the preferred direction in our setup).  Now, with the above form of the  wavenumber, the vector perturbations are represented by
\ba
\label{genericgaugefield}
\widehat{A}_{i}(\mathbf{k},\eta) = \sum _{\lambda=1,2}  \widehat{A}_{(\lambda) }(\mathbf{k},\eta)
\epsilon_{i}^{(\lambda)}(\bfk)  \quad , \quad 
\widehat{A}_{(\lambda) }(\mathbf{k},\eta) = 
  A_{(\lambda)}(k, \eta) a_{(\lambda)}(\mathbf{k})+ A^{*}_{(\lambda)}(k, \eta) a^{\dag}_{(\lambda)}(-\mathbf{k})    \, ,
\ea
in which  $ a_{(\lambda)}(\mathbf{k})$ and $a^{\dag}_{(\lambda)}(-\mathbf{k})$ are the annihilation and the creation operators associated with the gauge field perturbations  satisfying the commutation relation 
$ [a_{(\lambda)}(\mathbf{k}),   a^{\dag}_{(\lambda')}(\mathbf{k'})] = \delta_{\lambda \lambda'}\, \delta^{(3)} (\bfk-\bfk' )$. In addition,  $\epsilon_{i}^{(\lambda)}(\bfk) $ represents the gauge field transverse  polarization  satisfying  $k_{i}\epsilon_{i}^{(\lambda)}(\bfk) =0$. With the above choice for the wavenumber, the gauge field polarization is given by
\ba
\label{gaugefield polarization}
{\vec{\epsilon}}^{\,  \, (1)} = (\sin{\theta}, -\cos{\theta}, 0) \quad , \quad
\vec{\epsilon}^{\,  \, (2)} = (\cos{\theta}\cos{\varphi}, \sin{\theta}\cos{\varphi}, -\sin{\varphi})
\ea
In addition we define $A_{(1)}\equiv A$ and $ A_{(2)}\equiv B$. 

Similar to scalar field perturbation, in the absence of gravitational back-reactions,  the quantum fluctuations of the gauge field perturbations are given by
\ba
\label{gaugefield mode function}
 A_{k}(\eta)  = B_k(\eta) =
 \frac{i}{f \sqrt{2k}} \left( 1 - \frac{i}{k\eta} \right) e^{-ik\eta} \, .
\ea

Now we consider the tensor perturbations.  We decompose the tensor perturbations in terms of two polarization bases $e_{ij}^{(s)}(\bfk)$ with $s= \times, +$ satisfying
\ba
e^{(s)}_{ij}(\mathbf{k}) e^{*(s')}_{ij}(\mathbf{k}) = \delta_{ss'} \quad , \quad
e^{(s)}_{ij}(\mathbf{k}) = e^{*(s)}_{ij}(\mathbf{-k}) \, ,
\ea
in which a $*$ represents the complex-conjugation.
In addition,  the traceless and transverse conditions $h_{ii} = h_{ij,j} =0$, yields
\ba
e_{i i}^{(s)}(\bfk) = 0 \quad , \quad  k_j e_{i j}^{(s)}(\bfk) = 0  \, .
\ea

The quantum operators $\widehat{h}_{ij}(\mathbf{k},\eta) $ in Fourier space can be represented in terms of their annihilation and creation operators by
\ba
\label{tensor}
\widehat{h}_{ij}(\mathbf{k},\eta) = \sum _{s=+,\times}  \widehat{h}_{(s)}(\mathbf{k},\eta)
e_{ij}^{(s)}(\bfk)
 \quad , \quad \widehat{h}_{s}(\mathbf{k},\eta) \equiv
h_{(s)}(k, \eta)a_{s}(\mathbf{k})+ h^{*}_{(s)}(k, \eta)a^{\dag}_{s}(-\mathbf{k}) \, ,
\ea
with  the commutation relations  $ [ a_{s}(\bfk),    a_{s}^\dagger  (\bfk')] = \delta_{s s'} \delta^{(3)} (\bfk-\bfk' )$.

We also need to fix the tensor polarization. With the  wavenumber given by Eq. (\ref{k-general}),  the most general tensor polarization is given in Appen. \ref{generic tensor}. 
However, it turns out that for the calculation of the TSS correlation we still have a freedom to choose one of the fields in the $x-y$ plane. So in order to simplify the analysis, we can simply assume that the tensor mode is completely inside the  $x-y$ plane corresponding to  choosing  $\varphi =  \frac{\pi}{2} $ in the following. With this choice the polarizations 
$e^{+}_{ij}(\mathbf{k})$ and $e^{\times}_{ij}(\mathbf{k})$ become
\begin{align}
\label{polarization2+}
e^{+}_{ij}(\mathbf{k}) =&\frac{1}{\sqrt{2}} \left( \begin{array}{ccc}
 \sin^2{\theta}   & -\sin{\theta}\cos{\theta}& 0 \\
  -\sin{\theta}\cos{\theta} & \cos^2{\theta} & 0 \\
  0 & 0& -1 \\
\end{array} \right) \\
\label{polarization2times}
e^{\times}_{ij}(\mathbf{k}) =& \frac{1}{\sqrt{2}}\left( \begin{array}{ccc}
 0 & 0 & -\sin{\theta} \\
  0 & 0 & \cos{\theta} \\
  -\sin{\theta} & \cos{\theta} & 0 \\
\end{array} \right) ~.
\end{align}
As usual, the profile of the tensor excitations has the standard massless form
\ba
\label{hs-wave}
{h}_{(s)}(k,\eta) = \frac{2 i H\eta }{M_{P}\sqrt{2 k}}\left(1-\frac{i}{ k\eta} \right)e^{-ik\eta} ~~~,~~~ (s= +, \times) \, .
\ea
Correspondingly, the power spectrum of the tensor perturbations is given by
\ba
\langle   \widehat{h}_{ij}(\mathbf{k_1})  \widehat{h}_{ij}(\mathbf{k_2}) \rangle
= (2 \pi)^3 \delta^{(3)} (\bfk_1 + \bfk_2) P_h(k_1) \quad , \quad
\calP_h \equiv   \frac{ k_1^3  }{ 2 \pi^2}  P_h(k_1) \, .
\ea
In the isotropic case, the tensor power spectrum has the standard form
\ba
\label{calPh0}
\calP_h^{(0)} = \frac{2 H^2}{\pi^2 M_P^2}  = 16 \epsilon_H  \calP_\zeta^{(0)} \, .
\ea
Therefore, defining the tensor-to scalar ratio $r \equiv \calP_h/\calP_\zeta  $ we  obtain $r = 16 \epsilon_H$ for the isotropic theory.

\subsection{The anisotropic power spectrum}
\label{power}

As discussed before, the presence of a massless vector field interacting with the inflaton field induces 
anisotropy in scalar and tensor power spectrum which can be detected observationally.
All interactions responsible for generating anisotropic power spectra are pre-dominantly  
generated from the matter sector Lagrangian  \cite{Emami:2013bk}
\ba
\label{interaction Lagrangian}
L = -\frac{a^4}{4}f(\phi)^2 F_{\mu \nu} F^{\mu \nu} \, .
\ea
The anisotropic power spectrum is usually parameterized via \cite{Ackerman:2007nb}
\ba
\calP_\zeta = \calP_\zeta^{(0)}  \left( 1 + g_* \cos^2 \theta_{k}  \right)  \, ,
\ea
in which  $\theta_{k}$ is the angle between the preferred direction $\hat {\bf n}$  (in our example $\hat {\bf n} = \hat {\bf x}$) and the wavenumber $\bfk$ and  $g_*$ is the amplitude of quadrupole anisotropy.  
Using either the standard in-in formalism, as performed in  \cite{Watanabe:2010fh, Bartolo:2012sd, Emami:2013bk}, or the alternative $\delta N$ formalism  \cite{ Abolhasani:2013zya, Abolhasani:2013bpa}, the amplitude of quadrupole anisotropy in the attractor limit of \cite{Watanabe:2009ct}
is calculated to be 
\ba
\label{g-star-attractor}
g_* = -24 I N_e^2  \, .
\ea 
Using the Planck data, the bound $| g_*| \lesssim 10^{-2}$ is obtained in \cite{Kim:2013gka}. 
As a result, with $N_e \sim 60$ to solve the horizon and the flatness problem, one concludes that
$I \lesssim 10^{-7}$. Translating this to parameter $c$ appearing in gauge kinetic coupling, 
Eq. (\ref{f-form0}), one concludes that $c-1 \lesssim 10^{-7}$. On the other hand, in order for the system to reach the attractor regime during which the gauge field energy density is given by Eq. (\ref{R-attractor}), it is assumed in  \cite{Watanabe:2010fh} that $( c-1) N_e >1$. We clearly see that this condition is not  consistent with the observational bound on $g_*$ \cite{Naruko:2014bxa}. It is therefore necessary to revisit the anisotropy power spectrum without relying on the assumption of 
attractor  regime as performed in  \cite{Naruko:2014bxa}. In the limit where one neglects the slow-roll corrections 
and assuming the total number of e-folds $N_e$ are finite, the appropriate limit of the results in  
\cite{Naruko:2014bxa} yields 
\ba
\label{g-star-non-attractor}
g_*= - 24  \left( \frac{2 R}{\epsilon_H} \right) N_e^2 \, .
\ea
We have specifically checked that this result is also obtained using the $\delta N$ formalism
\cite{delta-N}. We note that  Eq. (\ref{g-star-non-attractor}) is consistent with the attractor limit
in which from Eq. (\ref{R-attractor}) one has $2 R/\epsilon_H = I$.  

In conclusion, as we have checked in \cite{delta-N}, 
the previous results of anisotropic curvature perturbation power spectrum and bispectrum obtained in 
\cite{Watanabe:2010fh, Bartolo:2012sd, Shiraishi:2013vja, Emami:2013bk,  Abolhasani:2013zya, Abolhasani:2013bpa,  Chen:2014eua, Shiraishi:2013oqa}
go through directly without relying on the attractor assumption 
upon replacing $I \rightarrow 2 R/\epsilon_H$. In particular, the amplitude of quadrupole anisotropy scales as $N_e^2$ which is supported from the accumulative  contributions of the super-horizon IR modes which have left the horizon and becomes classical \cite{ Bartolo:2012sd}.

\section{The TSS Bispectrum Correlations}
\label{interactions}

Having presented the free field mode functions, now we are in the position to consider the interaction Lagrangians and calculate the TSS bispectrum correlations.  As mentioned before, the leading interactions for anisotropies and the TSS correlations comes from the matter sector  Eq. (\ref{interaction Lagrangian}). 

Expanding the above action around the background values, one can obtain the interaction Lagrangian at the second and the third order levels. As usual, one can start by writing the interaction Lagrangian in the real space, as presented in Appendix \ref{int-lagrang}, and then try to transform it into the Fourier space. It turns out that the final result for our most general choice of wavenumber $\bfk$
is very complicated. While we present the full results in Appendix \ref{Fourierspacegeneric},  in what follows we only consider two simpler limits of the system. As we will see, these simple examples will be useful for us to better understand the leading order corrections.

\subsection{The all planar shapes for the bispectrum}
\label{example1}

Here we consider the simple case in which  all wavenumbers are co-planar extended in $x-y$ plane, corresponding to  $\varphi= \frac{\pi}{2}$. 

\subsubsection{{The second order interaction:}} 

In this case, the second order exchange vertex Lagrangians in the Fourier space have the following form,
\ba
\label{Lzetah}
L_{\zeta h_+} &=&
 - 6\sqrt{2} R M_{P}^2\int d^3k \sin^2{\theta_k}a^2\left(-\eta\right)^{-2} \zeta_\mathbf{k} {h}_{+\mathbf{k}} \\
\label{LzetaA}
 L_{\zeta A} &=& -4 M_{P}\sqrt{6 R}\int d^3k \sin{\theta_k} \left(\frac{a f}{\eta}\right) \zeta_\mathbf{k} A'_{\mathbf{k}} \\
 \label{Lh+D1}
 L_{h_+  A} &=& M_{P} \sqrt{3 R}
 \int d^3k \sin{\theta_k}\left( \frac{fa}{\eta}\right) A'_{\mathbf{k}} {h}_{+\mathbf{k}} \\
\label{LhxzD}
 L_{h_\times  B} &=& M_{P} \sqrt{3 R}\int d^3k \sin{\theta_k} \left(\frac{fa}{\eta}\right) B^{'}_\mathbf{k} \left({h}_{\times \mathbf{k}}\right)
\ea
where $\theta_{k}$ is the angle between the preferred direction $\hat {\bf n}$ and the wavenumber $\bf{k}$ (in our example $\hat {\bf n} = \hat {\bf x}$) and so on. 

\subsubsection{The Third order interaction:} 

Here we present the third order action. Since we would like to calculate the TSS bispectrum, we only present the leading contributions  into the TSS correlation. As we will see, at the leading order, there are five different diagrams which contribute into TSS correlation. They correspond to the following third order terms in the cubic action, $L^{SST}$, $L^{SSV}$, $L^{SVT}$, $L^{SVV}$ and $L^{VVT}$ where $S$, $V$ and $T$ respectively  refer to $\zeta$, $A$ and  $h_{+}$.

It is very important to note that beyond the above terms there are some other terms at the cubic action which turn out to be smaller than the leading corrections that we are interested in so throughout this work we do not consider them.  Fortunately there is an intuitive factor in order to determine the leading terms which we specify in what follows.  Moreover, there are also many other terms inside the above five terms which either do not appear in 
TSS or lead to a sub-leading contribution as compared to the leading term. We also neglect these terms and only mention the potentially leading contributions.

Now, before going ahead, it is worth to introduce the other terms inside the above five terms and mention briefly  why we are not interested in them. There are two types of terms in this regard:

\begin{itemize}

\item Terms which contain the spatial derivatives of the gauge field  corresponding to the magnetic field. As has been shown in the Appendix \ref{magnetic field suppression}, the final result for this part is logarithmically suppressed compared to the electric field coming from the  time derivative of the gauge field, so these terms do not contribute at the leading order. Intuitively, this comes from the fact that magnetic field decays outside the horizon while electric field is frozen on super-horizon scales.
 
\item Terms which either contain $B$ or $h_{\times}$.  However, in the planar case, $\varphi=\frac{\pi}{2}$, 
$\zeta$ only interacts with the scalar fields such as $A$ and $h_{+}$ at the second order level. As a result there is no leading sources for these types of interactions. 
\end{itemize}

Thanks to these simplifications for  the planar case, the third order interactions are simplified substantially given by 
\ba
\label{SST}
L^{SST} &=& -12 \sqrt{2} R M_{P}^2
\left(\frac{a^2}{\eta^2}\right) \int d^3p \int d^3q \sin^2{\theta_{(-p-q)}} \zeta_\mathbf{p} \zeta_\mathbf{q} h_{+(\mathbf{-p-q})} \\
\label{SSV}
L^{SSV} &=& -8 \sqrt{6 R} M_{P} \left(\frac{a f}{\eta}\right) \int d^3p \int d^3q \sin{\theta_{q}} \zeta_{(\mathbf{-p-q})} \zeta_\mathbf{p} A'_{\mathbf{q}} \\
\label{SVT}
L^{SVT} &=& 2 \sqrt{12 R} M_{P} \left(\frac{a f}{\eta}\right) \int d^3p \int d^3q \sin{\theta_{p}} \cos{(\theta_{p} - \theta_{q})} \zeta_{(\mathbf{-p-q})} h_{+\mathbf{p}} A'_{\mathbf{q}} \\
\label{SVV}
L^{SVV} &=& 2 f^2 \int d^3p \int d^3q \cos{(\theta_{p} - \theta_{q})} \zeta_{(\mathbf{-p-q})} A'_{\mathbf{p}} A'_{\mathbf{q}} \\
\label{VVT}
L^{VVT} &=& - \frac{\sqrt{2}}{4} f^2 \int d^3p \int d^3q \cos{(\theta_{p} - \theta_{(p+q)})} \cos{(\theta_{q} - \theta_{(p+q)})}
h_{+(\mathbf{-p-q})} A'_{\mathbf{p}} A'_{\mathbf{q}}
\ea

\subsection{TSS correlation function for the planar shape}

Having presented the cubic interactions, in what follows, by using the in-in formalism, we calculate the TSS bispectrum for each of the five mentioned pieces in the planar limit. As we shall see, this example turns out to be extremely useful for us in order to determine the leading term in the full TSS bispectrum. 

However, before going ahead, we emphasis that the standard in-in formalism is written in terms of  interaction Hamiltonian. So in order to use it we need to calculate the Hamiltonian from the interaction Lagrangian of the system. Moreover, in the presence of the kinetic interactions, $H_{int}$ is not simply $-L_{int}$. However, as it has been shown in \cite{Chen:2014eua},  the leading corrections in the correlation functions come entirely from the substitution $H_{int} =-L_{int}$ and we can safely use this prescription in the following analysis. 

\begin{figure}[!t]
	\centering
	\includegraphics[width=1.0\textwidth]{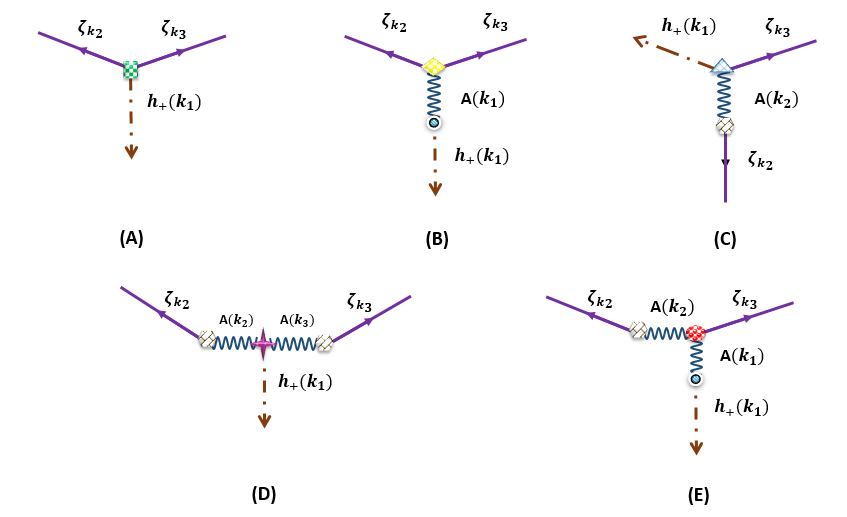}
	\caption{ Feynman diagrams for the Tensor-Scalar-Scalar cross-correlation with different types of exchange vertices.}
	\label{Feynmann}
\end{figure}

In order to have a better intuition about the potentially important terms it is worth to take a look at the Feynman diagrams in Fig. \ref{Feynmann}, representing how the above five terms can contribute into the TSS bispectrum.  From the plot we see that, $L^{SST}$, $L^{SSV}$, $L^{SVT}$, $L^{VVT}$   and $L^{SVV}$  correspond  respectively to the cases $A$, $B$, $C$, $D$ and $E$ in Fig. \ref{Feynmann}.

Now, before doing any analysis,  an important question is how can we determine  the potentially important contribution in the TSS bispectrum?  In order to find the answer, let us take a look at the simplest possible contribution in TSS which comes from $L^{SST}$ represented by diagram  $A$  in Fig. \ref{Feynmann}.  From Eq.  (\ref{SST}) we see that this interaction is proportional to $ R \ll 1$ which can be used  to determine  the potentially important terms in the TSS bispectrum. As we will show, all  terms in Fig. \ref{Feynmann}
are proportional to $R$.

The next question, related to the above question,   is whether all diagrams  presented in 
Fig. \ref{Feynmann} are important in performing the in-in analysis. Actually, it turns out that not all of these terms are the same order. After performing the time integrals  the leading contributions come from $L^{SVV}$ corresponding to diagram $E$ in  Fig. \ref{Feynmann}. Having this said, in the following, we calculate TSS bispectrum for all diagrams in Fig. \ref{Feynmann}.  Here we only present the final result for each diagram and leave the detail analysis to Appendix. \ref{in-in}.


\begin{itemize}

\item \textbf{SST}: This corresponds  to diagram $A$ in  Fig. \ref{Feynmann} for which we need to calculate in-in integral to first order.  The result is obtained to be 
\ba
\left\langle h_{+\mathbf{k_1}} \zeta_{\mathbf{k_2}} \zeta_{\mathbf{k_3}}(\eta_e) \right\rangle = -\left(\frac{\sqrt{2} R}{\epsilon_H}  N_e\right)
\left[ \sin^2{\theta_{k_1}} \left(\frac{k_1^3 + k_2^3 + k_3^3}{k_3^3}\right) P_{h_{+\, \mathbf{k_1} }}  P_{\zeta \, \mathbf{k_2}} + (\mathbf{k_2} \leftrightarrow \mathbf{k_3})\right] \, ,
\ea
where we have defined the number of e-folds as $N_e \equiv - \ln{\left(\frac{\eta_e}{\eta_0}\right)} $
and $\theta_{k_1}$ is the angle between the preferred direction $\hat {\bf n}$ and the wavenumber $\bf{k_1}$ (in our example $\hat {\bf n} = \hat {\bf x}$).

\item \textbf{SSV}: This term corresponds to diagram  $B$ in  Fig. \ref{Feynmann}. In this case, we have to calculate the in-in integral to second order. The final result is obtained to be 
\ba
\left\langle h_{+\mathbf{k_1}} \zeta_{\mathbf{k_2}} \zeta_{\mathbf{k_3}} (\eta_e)\right\rangle = - 
\left(\frac{2 \sqrt{2} R}{\epsilon_H}\right)
\left(3 N_e^2 - 2 N_e \right)
 \sin^2{\theta_{k_1}}
 \left[ \left(\frac{k_2^3 + k_3^3}{k_3^3}\right)
 P_{\zeta\, \mathbf{k_2}} P_{h_{+\,  \mathbf{k_1} }} + (\mathbf{k_2} \leftrightarrow \mathbf{k_3})\right] \, .
\ea

\item \textbf{SVT}: This term corresponds to diagram $C$ in Fig. \ref{Feynmann}. In this case, we have to calculate  the in-in integral to second order, yielding 
\ba
\left\langle h_{+\mathbf{k_1}} \zeta_{\mathbf{k_2}} \zeta_{\mathbf{k_3}} (\eta_e)\right\rangle &=& -
\left(\frac{4 \sqrt{2} R}{\epsilon_H}\right)
\left(3 N_e^2 - 2 N_e \right)   \nonumber \\
&\times&\bigg{[} \sin{\theta_{k_1}}\sin{\theta_{k_2}} \cos{(\theta_{k_1}- \theta_{k_2} )}\left(\frac{k_1^3 + k_3^3}{k_3^3}\right)
 P_{\zeta \mathbf{k_2}}P_{h_{+\,  \mathbf{k_1}} } + (\mathbf{k_2} \leftrightarrow \mathbf{k_3})\bigg{]}\, . \nonumber
\ea

\item \textbf{VVT}: This term corresponds  to diagram $D$ in Fig. \ref{Feynmann}. For this case, one has to expand the in-in integral up to third order in the perturbations yielding 
\ba
\left\langle h_{+\mathbf{k_1}} \zeta_{\mathbf{k_2}} \zeta_{\mathbf{k_3}} (\eta_e)
\right\rangle &=& - 
\left(\frac{18 \sqrt{2} R}{\epsilon_H}\right)
\left( N_e^3 - N_e^2  + \frac{1}{3} N_e \right)  \left(\frac{k_1^3}{k_3^3}\right)  \nonumber \\
&\times&
\bigg{[} \sin{\theta_{k_2}}\sin{\theta_{k_3}} \cos{(\theta_{k_1}- \theta_{k_2} )} \cos{(\theta_{k_1}- \theta_{k_3} )} P_{\zeta\mathbf{k_2}}P_{h_+ \mathbf{k_1}}\bigg{]} + (\mathbf{k_2} \leftrightarrow \mathbf{k_3}) \, .
\ea
\item \textbf{SVV}: This term corresponds to diagram $E$ in Fig. \ref{Feynmann}. Like in the previous case, one has to calculate the in-in integral to  third order yielding 
\ba
\left\langle h_{+\mathbf{k_1}} \zeta_{\mathbf{k_2}} \zeta_{\mathbf{k_3}} (\eta_e)\right\rangle &=& -
\left(\frac{36 \sqrt{2} R}{\epsilon_H}\right)
\left( N_e^3 - N_e^2 +\frac{1}{3} N_e \right)   \nonumber \\
&&
\times \bigg{[ } \sin{\theta_{k_1}}\sin{\theta_{k_2}} \cos{(\theta_{k_1}- \theta_{k_2} )}
 P_{\zeta}(\mathbf{k_2})P_{h_+}(\mathbf{k_1}) + (\mathbf{k_2} \leftrightarrow \mathbf{k_3})\bigg{] } \, .
\ea
\end{itemize}

Now  adding the above five contributions the final result is obtained to be 
\ba
\left\langle h_{+\mathbf{k_1}} \zeta_{\mathbf{k_2}} \zeta_{\mathbf{k_3}} (\eta_e)\right\rangle&= &
\left\langle h_{+\mathbf{k_1}} \zeta_{\mathbf{k_2}} \zeta_{\mathbf{k_3}} (\eta_e)\right\rangle_{A} +
\left\langle h_{+\mathbf{k_1}} \zeta_{\mathbf{k_2}} \zeta_{\mathbf{k_3}} (\eta_e)\right\rangle_{B} + \left\langle h_{+\mathbf{k_1}} \zeta_{\mathbf{k_2}} \zeta_{\mathbf{k_3}} (\eta_e)\right\rangle_{C} \nonumber\\
&+& \left\langle h_{+\mathbf{k_1}} \zeta_{\mathbf{k_2}} \zeta_{\mathbf{k_3}} (\eta_e)\right\rangle_{D} + \left\langle h_{+\mathbf{k_1}} \zeta_{\mathbf{k_2}} \zeta_{\mathbf{k_3}} (\eta_e)\right\rangle_{E} \nonumber\\
& \simeq& - 
\left(\frac{36 \sqrt{2} N_e^3 R}{\epsilon_H}\right)
\bigg{[} \sin{\theta_{k_1}}\sin{\theta_{k_2}} \cos{(\theta_{k_1}- \theta_{k_2} )}
P_{\zeta}(\mathbf{k_2})P_{h_+}(\mathbf{k_1}) + (\mathbf{k_2} \Leftrightarrow \mathbf{k_3})\bigg{]} \nonumber\\
& \simeq & \left(\frac{3 g_* N_e }{2}\right) \left( \left[ \mathbf{\hat{x}}\cdot e^{+}(\mathbf{k_1})\cdot \mathbf{\hat{x}} - (\mathbf{\hat{k}_3}\cdot \mathbf{\hat{x}}) \left(\mathbf{\hat{x}}\cdot e^{+}(\mathbf{k_1})\cdot \mathbf{\hat{k}_3} \right) \right] P_{\zeta\mathbf{k_3}} P_{ h_{+ \mathbf{k_1}}} + 
  \mathbf{k_3} \leftrightarrow \mathbf{k_2} \right) \, , 
\ea
where we have only considered the leading contribution in the final result which comes from SVV 
represented by diagram $E$ in  Fig. \ref{Feynmann}.


\subsection{Planar tensor and non-planar vector fields}
\label{example2}

As a second example, in the following, we  consider the case where the tensor field is in the $x-y$ plane, i.e. $\varphi_{\mathrm{tensor}} = \frac{\pi}{2}$, while the polarization of vector field has the general non-planar form. In this case, the situation is more complicated than the first example studied above  and  the above 5 diagrams are converted into 13 diagrams. As a result the situation is much more complicated than the previous  example in which all modes were assumed to be coplanar. However, one does not need to worry about these complexities. This is because, as we have already proved in the previous example, there is a hierarchy between different terms and thanks to this hierarchy, we only need to consider the last diagram, the diagram $E$ in Fig. \ref{Feynmann}. Since this diagram comes from the SVV vertex, in the following, we only need to consider this vertex and do not consider the other 12 diagrams. Having this said, we only consider the second and third order interactions   relevant  for this diagram.

\subsubsection{The Second order interaction:} 

The relevant second ordered interactions are given by,
\ba
\label{LSV}
L_{SV} &=& -4 M_{P}\sqrt{2 R}\int d^3k \left(\frac{a f}{\eta}\right) \zeta_{\mathbf{-k}} \left(\sin{\theta_{k}} A'_{\mathbf{k}} + \cos{\theta_{k}}\cos{\varphi_{k}} B'_{\mathbf{k}} \right) \\
\label{LVT}
L_{VT} &=& M_{P} \sqrt{3 R}
\int d^3k \left( \frac{af}{\eta}\right) \left(\sin{\theta_{k}}A'_{\mathbf{-k}} {h}_{+\mathbf{k}} + \sin{\theta_{k}}B'_{\mathbf{-k}} {h}_{\times \mathbf{k}} \right) \, .
\ea
\\

\subsubsection{\textit{The Third order interaction:}}

The third order interactions is given by,
\ba
\label{LSVV}
L_{SVV} = 2 f^2 \int d^3p d^3q  \zeta_{(\mathbf{-p-q})} 
\bigg{[} \cos{(\theta_p -\theta_q )} A'_{\mathbf{p}} A'_{\mathbf{q}} + 2 \sin{(\theta_p -\theta_q)} \cos{\varphi_{p}} A'_{\mathbf{q}} B'_{\mathbf{p}} + \sin{\varphi_{p}} \sin{\varphi_{q}} B'_{\mathbf{p}} B'_{\mathbf{q}} 
\bigg{]} \, .
\ea

\subsection{TSS correlation for the planar tensor and general non-planar vector fields}

\begin{figure}[!h]
	\centering
	\includegraphics[width=1.0\textwidth]{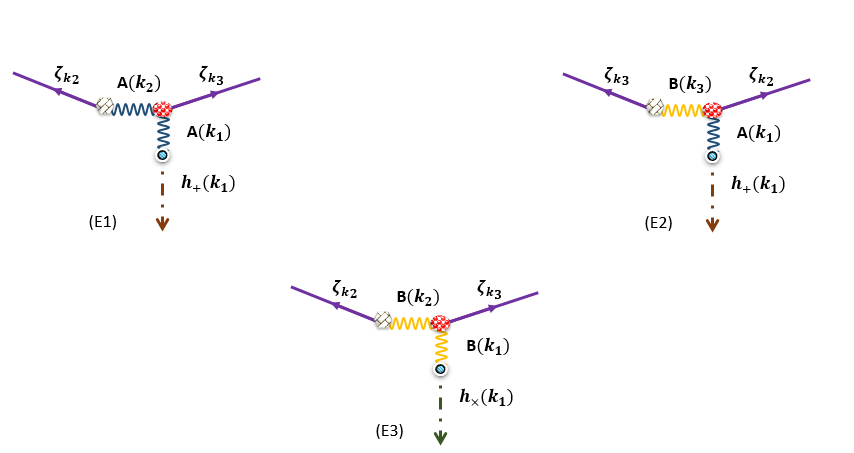}
	\caption{Leading order corrections on the TSS cross-correlation function for the non-planar case
	\label{PlanarT}.}
\end{figure}

Here we  calculate the TSS correlation function for the planar tensor field with the  vector field having 
the general non-planar wavenumber. 
As we mentioned  above, we consider only the most relevant diagram for this case which comes from the SVV type interaction. As it has been shown in  Fig. \ref{PlanarT}, in this case there are three different contributions that are worth to be considered. \\

$\blacklozenge$ First we calculate  $\left\langle h_{+k_1} \zeta_{k_2} \zeta_{k_3} (\eta_e)\right\rangle$ which is 
\ba
\label{TSSfinal1}
\left\langle h_{+\mathbf{k_1}} \zeta_{\mathbf{k_2}} \zeta_{\mathbf{k_3}} (\eta_e)\right\rangle &=&  -
\left(\frac{36 \sqrt{2} R}{\epsilon_H}\right)
\left( N_e^3 - N_e^2 +\frac{ N_e}{3} \right) \nonumber\\ 
&\times&\bigg{[} \left(\sin{\theta_{k_1}}\sin{\theta_{k_3}} \cos{(\theta_{k_1}- \theta_{k_3} )} +  \sin{\theta_{k_1}} \cos{\theta_{k_3}} \cos^2{\varphi_{k_3}} \sin{(\theta_{k_3}- \theta_{k_1} )}\right) 
P_{\zeta \mathbf{k_3}} P_{h_+ \mathbf{k_1}} +
(\mathbf{k_3} \leftrightarrow \mathbf{k_2})\bigg{]}
\nonumber\\
&\simeq&  \frac{3}{2} g_{*} N_e \bigg{[}\left(\mathbf{\hat{x}}\cdot e^{+}(\mathbf{k_1})\cdot \mathbf{\hat{x}} - (\mathbf{\hat{k}_3}\cdot \mathbf{\hat{x}}) (\mathbf{\hat{x}}\cdot e^{+}(\mathbf{k_1})\cdot \mathbf{\hat{k}_3}) \right) P_{\zeta \mathbf{k_3}} P_{ h_{+ \mathbf{k_1}}} + (\mathbf{k_3} \leftrightarrow \mathbf{k_2})\bigg{]}  \, ,
\ea
where,  as before, $ \mathbf{\hat{x}} $ refers to the unit vector in the preferred direction and we have used Eq. (\ref{g-star-non-attractor}) to express the final result in terms of
the amplitude of quadrupole anisotropy $g_*$. \\

$\blacklozenge$ Now we calculate  $\left\langle h_{\times k_1} \zeta_{k_2} \zeta_{k_3} (\eta_e)\right\rangle$, obtaining 
\ba
\label{TSSfinal2}
\left\langle h_{\times \mathbf{k_1}} \zeta_{\mathbf{k_2}} \zeta_{\mathbf{k_3}} (\eta_e)\right\rangle &=& - \left(\frac{36 \sqrt{2} R}{\epsilon_H}\right)
\left( N_e^3 - N_e^2 +\frac{1}{3} N_e \right)
\bigg{[} \sin{\theta_{k_1}}\cos{\theta_{k_3}} \sin{\varphi_{k_3}}\cos{\varphi_{k_3}}
P_{\zeta \mathbf{k_3}} P_{h_+ \mathbf{k_1}} +
(\mathbf{k_3} \leftrightarrow \mathbf{k_2})\bigg{]}\nonumber\\&&
\sim \frac{3}{2} g_{*} N_e \bigg{[}\left(\mathbf{\hat{x}}\cdot e^{\times}(\mathbf{k_1})\cdot \mathbf{\hat{x}} - (\mathbf{\hat{k}_3}\cdot \mathbf{\hat{x}}) (\mathbf{\hat{x}}\cdot e^{\times}(\mathbf{k_1})\cdot \mathbf{\hat{k}_3}) \right) P_{\zeta}(\mathbf{k_3}) P_{ h_{\times}}(\mathbf{k_1}) + (\mathbf{k_3} \leftrightarrow \mathbf{k_2})\bigg{]}  \, .
\ea

There are two non-trivial features in these TSS correlation functions that in the following we  elaborate in some details. 

$\bullet$ First of all, we emphasis that there are two different sources for the TSS correlations. One comes from the isotropic part of the action, which is similar to the single field inflation and is almost canceled by the other terms. The other term which we are interested in here is indeed due to the anisotropic matter corrections to the FRW universe which is determined  by 
the amplitude of the quadrupole anisotropy in power spectrum  $g_{*} = - \left(\frac{48 R}{\epsilon_H}\right) N_e^2$. One feature of this correction is that  it has primordial IR  origin and  it can not be canceled by any late time effects such as the non-linear effect as well as the projection effect. Yet, this correction is under control as it is proportional to the quadrupole anisotropy at the two point level.

$\bullet$ The second interesting point is that the shape of the TSS is not the same as in the well known fossil shape given in \cite{Jeong:2012df}. This is because in our model, in addition to the 
wavenumbers and the polarization tensor, we also have the effect of preferred direction, i.e. the statistical anisotropy. As a result  one expects that this preferred direction would  show up in the final results. 
 
\section{Observational Signatures}
\label{observation}

In this Section, we present different observational signatures of our anisotropic set up.

\subsection{SS vs TSS power asymmetry}
As we have seen  breaking the statistical isotropy leads to a quadrupole anisotropy at the two point level of the primordial curvature perturbation  with the following form,
\ba
\label{powerasymmetrytwopoint}
P_{\zeta}(\mathbf{K_S}) = P^{0}_{\zeta} \left( 1 + g_{*} (\mathbf{\hat{K}_S}\cdot \mathbf{\hat{x}})^2 \right) \, ,
\ea
with $g_{*} = -\left(\frac{48 R}{\epsilon_H}\right) N_e^2$.  Here we have shown that in addition to the above quadrupole anisotropy in the power spectrum, there is also another prediction from anisotropic inflation which comes from the anisotropic correlation between the short scalar  modes and long tensor modes,
\ba
\label{powerasymmetrythreepoint}
P_{\zeta}(\mathbf{K_S})\bigg{|}_{h_{(\lambda)}(\mathbf{K_L})} = P_{\zeta}(k_S) \Delta^{ij}h_{ij}^{(\lambda)}(\mathbf{K_L}) \, ,
\ea
where $\Delta_{ij}$ is given by, 
\ba
\Delta^{ij} &=& -
\left(\frac{144 R N_e^3}{\epsilon_H}\right)
 \bigg{(}\mathbf{\hat{x}^{i}} \mathbf{\hat{x}^{j}} - (\mathbf{\hat{K}_S}\cdot \mathbf{\hat{x}})(\mathbf{\hat{x}^{i}}\mathbf{\hat{K}_S^{j}}) \bigg{)} \nonumber\\
&& = 3 g_{*} N_e \bigg{(}\mathbf{\hat{x}^{i}} \mathbf{\hat{x}^{j}} - (\mathbf{\hat{K}_S}\cdot \mathbf{\hat{x}})(\mathbf{\hat{x}^{i}}\mathbf{\hat{K}_S^{j}}) \bigg{)}
\ea
So before going ahead, it is suitable to compare the results from the two-point level with the result at the cubic order. We can compare the amplitude as well as the shape of these two functions with each other.  Let us first compare the amplitudes. 
The ratio between the amplitudes of the power asymmetry at the two point as well as the three point levels is given by, 
\ba
\frac{\delta\left(P_{\zeta}(\mathbf{K_S})|_{h_{(\lambda)}(\mathbf{K_L})} \right) }{\delta \left(P_{\zeta}(\mathbf{K_S})\right)} = 3 N_{e} h_{(\lambda)}(\mathbf{K_L}) \, .
\ea
For reasonable values of total number of e-foldings, say  $N_e < 100$, the above ratio would be of order $10^{-3}$. This means that the quadratic corrections are more important than the cubic corrections in this set up. 

As for the shapes,  comparing Eq. (\ref{powerasymmetrytwopoint}) with Eq. (\ref{powerasymmetrythreepoint}), we see that the shapes of these asymmetries are not the same. So in principle one can distinguish them from each other.  


\subsection{Quadrupole Anisotropy in the Galaxy Power Spectrum}

Here we  find the shape of the galaxy power asymmetry within our anisotropic setup. In order to do this, we would use the orthogonality of the spherical harmonics, $Y_{LM}$. In addition, it is clear that the asymmetry is in the form of  quadrupole which means that we are only interested in the $L=2$ case. So we would have,
\ba
\label{multipoe moments}
Q_{2M} &=& \frac{\int d^2\hat{k}P_{\zeta}(|k|)\Delta^{ij}h^{(\lambda)}_{ij}(\mathbf{K_L})Y^{*}_{2M}}{\int d^2\hat{k}P_{\zeta}(|k|)Y^{*}_{00}} \nonumber\\
&=&\int d^2\hat{k}P_{\zeta}(|k|)\Delta^{ij}h^{(\lambda)}_{ij}(\mathbf{K_L})Y^{*}_{2M}
\ea
We note that the quadrupole components are not a good observable by themselves and in order to get the observables one needs to calculate the root-mean-square of the above quantity which is given by, 
\ba
\label{root-mean-square}
\overline{Q}^2 \equiv \bigg{\langle} \sum_{M=-2}^{+2} | Q_{2M}|^2 \bigg{\rangle} \, .
\ea
In addition, as we shall see below, depending on our purpose for the detection of the tensor mode, we may also need averaging over the long mode and in the case that we do not have any information about the preferred direction, we also perform an averaging over this direction. Keeping these in mind, in the remaining part of this section, we try to calculate the root-mean-square of the quadrupole moment. As soon as we calculate it, depending on our purpose, we may also do the averaging over the long-mode as well as the preferred direction.

After an extensive analysis, which is presented in the App. (\ref{quadrupole rms}), the root-mean-square of the quadrupole is obtained to be
\ba
\label{root-mean-square final}
\overline{Q}^2 &=& (8\pi) \left(g_{*}N_{e}\right)^2 P_{h}(K_{L}) \left(1 - (\mathbf{\hat{K}_{L}}\cdot \mathbf{\hat{x}})^2\right)\bigg{(} 1 - \left(40 \pi^2\right) \sum_{s,s',q,q' =-1}^{+1} \sum_{t=-2}^{+2} Y_{1s}^{*}(\mathbf{\hat{x}}) Y_{1s'}^{*}(\mathbf{\hat{x}}) Y_{1q}^{*}(\mathbf{\hat{K}_{L}}) Y_{1q'}^{*}(\mathbf{\hat{K}_{L}}) \nonumber\\
&&\left( \begin{array}{ccc}
	2 & 1 & 1\\
	0 & 0 & 0
\end{array} \right)^2
\left( \begin{array}{ccc}
	2 & 1 & 1\\
	t & q & s
\end{array} \right)
\left( \begin{array}{ccc}
	2 & 1 & 1\\
	-t & q' & s'
\end{array} \right) 
\bigg{)} \, ,
\ea
where $\left( \begin{array}{ccc}
l & l_1 & l_2\\
m & m_1 & m_2
\end{array} \right)$ is the Wigner 3-j symbol.\\

\textbf{{ \textit{Averaging over the long mode :}} }\\

In the next step we  calculate the average of the root-mean-square of the quadrupole anisotropy over the long mode as follows
\ba
\label{rms-long mode average}
\overline{Q}^2|_{L} &\equiv& \int _{K_{Lmin}}^{K_{Smin}} d^3 K_{L} \overline{Q}^2 (\mathbf{K_{L}}, \mathbf{\hat{x}}) \nonumber\\
&=& (8\pi) \left(g_{*}N_{e}\right)^2 \left(\frac{H}{M_P}\right)^2 \ln{\left(\frac{K_{Smin}}{K_{Lmin}}\right)} \int d\Omega_{\mathbf{K_{L}}} \left(1 - (\mathbf{\hat{K}_{L}}\cdot \mathbf{\hat{x}})^2\right)\bigg{(} 1 - \left(40 \pi^2\right) \nonumber\\
&& \sum_{s,s',q,q' =-1}^{+1} \sum_{t=-2}^{+2} Y_{1s}^{*}(\mathbf{\hat{x}}) Y_{1s'}^{*}(\mathbf{\hat{x}}) Y_{1q}^{*}(\mathbf{\hat{K}_{L}}) Y_{1q'}^{*}(\mathbf{\hat{K}_{L}})
\left( \begin{array}{ccc}
	2 & 1 & 1\\
	0 & 0 & 0
\end{array} \right)^2
\left( \begin{array}{ccc}
	2 & 1 & 1\\
	t & q & s
\end{array} \right)
\left( \begin{array}{ccc}
	2 & 1 & 1\\
	-t & q' & s'
\end{array} \right)
\bigg{)} \nonumber\\
&=& (8\pi) \left(g_{*}N_{e}\right)^2 \left(\frac{H}{M_P}\right)^2 \ln{\left(\frac{K_{Smin}}{K_{Lmin}}\right)} \bigg{(} \frac{8\pi}{3} - \left(\frac{80\pi^2}{3}\right)  \sum_{s,s',q =-1}^{+1} \sum_{t=-2}^{+2} (-1)^q Y_{1s}^{*}(\mathbf{\hat{x}}) Y_{1s'}^{*}(\mathbf{\hat{x}}) \nonumber\\
&&
 \left( \begin{array}{ccc}
	2 & 1 & 1\\
	0 & 0 & 0
\end{array} \right)^2
\left( \begin{array}{ccc}
	2 & 1 & 1\\
	t & q & s
\end{array} \right)
\left( \begin{array}{ccc}
	2 & 1 & 1\\
	-t & -q & s'
\end{array} \right) + 
\left(\frac{64\pi^3}{3}\right) \sqrt{\frac{45}{4\pi}} \sum_{s,s',q,q' =-1}^{+1} \sum_{t=-2}^{+2} \sum_{m=-2}^{+2} (-1)^{q+q'} Y_{1s}^{*}(\mathbf{\hat{x}}) Y_{1s'}^{*}(\mathbf{\hat{x}}) Y_{2m}^{*}(\mathbf{\hat{x}}) \nonumber\\
&&
\left( \begin{array}{ccc}
	2 & 1 & 1\\
	0 & 0 & 0
\end{array} \right)^3
\left( \begin{array}{ccc}
	2 & 1 & 1\\
	t & q & s
\end{array} \right)
\left( \begin{array}{ccc}
	2 & 1 & 1\\
	-t & q' & s'
\end{array} \right)
\left( \begin{array}{ccc}
	2 & 1 & 1\\
	m & -q' & -q
\end{array} \right)
\bigg{)}
\ea
Interestingly we see that in this model the averaged power quadrupole anisotropy is logarithmically sensitive to the number of inflationary modes in the survey. \\

\textbf{\textit{Averaging over the preferred direction:}} \\

So far we had assumed that we know the preferred direction in the sky (the direction of anisotropy) from other experiments. Now one can ask what will be our observables when there is no knowledge about the direction of anisotropy. In order to calculate the observed quadrupole anisotropy in this case one needs to average over the preferred directions. So in the following we  also calculate the averaged power quadrupole over different preferred directions. We start with Eq. (\ref{rms-long mode average}) and perform averaging over different values of the preferred direction
\ba
\label{rms-preferred direction averaged}
\overline{Q}^2|_{F} &\equiv& \int d^2\hat{x} \overline{Q}^2|_{L}(\mathbf{\hat{x}})\nonumber\\
&=& (8\pi) \left(g_{*}N_{e}\right)^2 \left(\frac{H}{M_P}\right)^2 \ln{\left(\frac{K_{Smin}}{K_{Lmin}}\right)} \bigg{(} \frac{32\pi^2}{3} - \left(\frac{80\pi^2}{3}\right)  \sum_{s,q =-1}^{+1} \sum_{t=-2}^{+2} (-1)^{q+s} \nonumber\\
&&
\left( \begin{array}{ccc}
	2 & 1 & 1\\
	0 & 0 & 0
\end{array} \right)^2
\left( \begin{array}{ccc}
	2 & 1 & 1\\
	t & q & s
\end{array} \right)^2
 + 
\left(240\pi^2\right) \sum_{s,s',q,q' =-1}^{+1} \sum_{t,m=-2}^{+2} (-1)^{q+q'}  \nonumber\\
&&
\left( \begin{array}{ccc}
	2 & 1 & 1\\
	0 & 0 & 0
\end{array} \right)^4
\left( \begin{array}{ccc}
	2 & 1 & 1\\
	t & q & s
\end{array} \right)
\left( \begin{array}{ccc}
	2 & 1 & 1\\
	-t & q' & s'
\end{array} \right)
\left( \begin{array}{ccc}
	2 & 1 & 1\\
	m & -q' & -q
\end{array} \right)
\left( \begin{array}{ccc}
	2 & 1 & 1\\
	-m & -s' & -s
\end{array} \right)
\bigg{)}
\ea


\subsection{Optimal Estimator for the Anisotropic Tensor mode}

Since the TSS correlation function is proportional to the tensor amplitude, one can use this set up to estimate the tensor modes within the anisotropic inflationary models.  In the following, we  try to do this in our set up. We start by writing down the minimum variance estimator for the tensor mode and then will continue with calculating the optimal estimator. Then we discuss briefly about the observability of the tensor modes in this set up by using the current and future galaxy surveys.

As it has been mentioned in \cite{Jeong:2012df}, in the presence of the tensor-scalar-scalar correlation, each pair of  $\delta(\mathbf{k_2})$ and $\delta(\mathbf{k_3})$  provide an estimator for the tensor mode with $\mathbf{K} = \mathbf{k_2} + \mathbf{k_3}$,  where we have redefined $(\mathbf{-k_1})$ as $\mathbf{K}$. 
\ba
\widehat{h_{(\lambda)}(\mathbf{K})} = \frac{\delta(\mathbf{k_2})\delta(\mathbf{k_3})}{P(k_2)}\left(\Delta^{ij}(\mathbf{\hat{x}},\mathbf{\hat{k_2}}) e^{(\lambda)}_{ij}(\mathbf{K})\right)^{-1}
\ea
In the following, we assume the null hypothesis, in which in the absence of the tensor mode the density perturbations are statistically isotropic and Gaussian. Of course in our current set up it is not completely true because as we have mentioned above we have broken the statistical isotropy so we have some anisotropic corrections in the power spectrum of the density perturbations. However, the level of anisotropy is very small and one can neglect them at this level. Naively, it  means that we can write  $\langle\delta(\mathbf{k})\delta(\mathbf{k'})\rangle = V \delta^{D}_{\mathbf{k},\mathbf{-k'}} P^{tot}(k)$, where $V$ appears here due to the fact that we have discretized the density function and $\delta^{D}_{\mathbf{k},\mathbf{-k'}}$ is the Kronecker delta function and $P^{tot} = P(\mathbf{k}) + P^{(n)}(\mathbf{k})$ is the measured matter power spectrum, which is including both the signal, $P(\mathbf{k})$, and the noise, $P^{(n)}(\mathbf{k})$. 

We can then calculate the variance of the above estimator, 
\ba
\mathrm{Variance} = 2 V P^{tot}(\mathbf{k_2}) P^{tot}(\mathbf{k_3}) \bigg{|}\Xi^{ij}(\mathbf{\hat{x}},\mathbf{\hat{k}_2}) e^{(\lambda)}_{ij}(\mathbf{K})\bigg{|}^{-2} \, ,
\ea
where the factor  2 comes from the permutation and $\Xi^{ij}(\mathbf{\hat{x}},\mathbf{\hat{k}_2}) = \Delta^{ij}(\mathbf{\hat{x}},\mathbf{\hat{k}_2})P(\mathbf{k}_2)$. 

The \textit{minimum-varince estimator} for the $h_{(\lambda)}(\mathbf{K})$ can then be calculated by summing over the whole individual pairs and with the inverse-variance weighting, 
\ba
\label{minimumes}
\widetilde{h_{(\lambda)}(\mathbf{K})} = P^{(n)}_{(\lambda)}(\mathbf{K}) \sum_{\mathbf{k}}\frac{\Xi^{ij}(\mathbf{\hat{x}},\mathbf{\hat{k}}, \mathbf{\hat{K}-\hat{k}}) e^{(\lambda)}_{ij}(\mathbf{K})}{2V P^{tot}(\mathbf{k})P^{tot}(\mathbf{K-k})}\delta(\mathbf{k_2})\delta(\mathbf{k_3}) \, ,
\ea
where in order to emphasis the symmetric behavior  under $\mathbf{k_2} \leftrightarrow \mathbf{k_3}$, we have presented this function more precisely as $\Xi^{ij}(\mathbf{\hat{x}},\mathbf{\hat{k}}, \mathbf{\hat{K}-\hat{k}})$. The reason is that, from Eqs.(\ref{TSSfinal1}) and (\ref{TSSfinal2}), it is clear that due to the permutation between $\mathbf{k_2}$ and $\mathbf{k_3}$ the answer should be symmetric. So far we did not take this into account  because we are in the squeezed limit and these two modes are equal to each other at the leading order. However, it is important to keep in mind that the precise notation is the one given by Eq. (\ref{minimumes}).

In addition the noise power spectrum is also given by 
\ba
\label{noisepower}
P^{(n)}_{(\lambda)}(\mathbf{K}) = \Bigg{[} \sum_{\mathbf{k}} \frac{\bigg{|}\Xi^{ij}(\mathbf{\hat{x}},\mathbf{\hat{k}}, \mathbf{\hat{K}-\hat{k}}) e^{(\lambda)}_{ij}(\mathbf{K})\bigg{|}^2}{2V P^{tot}(k)P^{tot}(|K-k|)} \Bigg{]}^{-1} \, .
\ea
In the most general case, the tensor mode, $h_{(\lambda)}(\mathbf{K})$ arises as the realizations of the random fields with the power spectrum, $P_{(\lambda)}(\mathbf{K}) = A_{(\lambda)} P^f_{(\lambda)}(\mathbf{K})$, where $\lambda$ denotes the polarizations of the tensor mode while $A_{(\lambda)}$ refers to the amplitude of the tensor mode and $P^f_{(\lambda)}(\mathbf{K})$ is the fiducial power spectrum. 

Now each Fourier mode of the $h_{(\lambda)}(\mathbf{K)}$ provides an estimator for the amplitude of the tensor field as follows
\ba
\label{Amplitudeestimator}
\widehat{A_{(\lambda)}(\mathbf{K})} = \bigg{[} P^f_{(\lambda)}(\mathbf{K}) \bigg{]}^{-1} \left(V^{-1} \bigg{|} \widehat{h_{(\lambda)}(\mathbf{K})}\bigg{|} - P^{n}_{(\lambda)}(\mathbf{K}) \right) \, ,
\ea
where we have subtracted the noise contribution to unbias the estimator. Now, as we mentioned above, each pair of the density fields would act as an estimator for the gravitational wave. So if we have a huge number of the density fields then according to the central limit theorem
$h_{(\lambda)}(\mathbf{K})$ would be very close to a Gaussian variable.  As a result the variance of the estimator in Eq. (\ref{Amplitudeestimator}), under the null hypothesis,  would be
\ba
\mathrm{Variance~of~Amplitude} = 2 \bigg{[} P^f_{(\lambda)}(\mathbf{K}) \bigg{]}^{-2} \bigg{[} P^{(n)}_{(\lambda)}(\mathbf{K}) \bigg{]}^{2} \, .
\ea
Now adding the estimators for every Fourier modes with inverse variance weighting yields the \textit{optimal estimator},
\ba
\label{optimales}
\widetilde{A_{(\lambda)}(\mathbf{K})} = \sigma_{h}^2\sum_{\mathbf{K},\lambda} \frac{ P^f_{(\lambda)}(\mathbf{K})}{2\bigg{[} P^{n}_{(\lambda)}(\mathbf{K}) \bigg{]}^{2}} \left( V^{-1} \bigg{|} \widehat{h_{(\lambda)}(\mathbf{K})}\bigg{|}^2 - P^{n}_{(\lambda)}(\mathbf{K})\right) \, ,
\ea
where we have 
\ba
\label{optimalvar}
\sigma_{h}^{-2} = \sum_{K,\lambda} \frac{\bigg{[} P^f_{(\lambda)}(\mathbf{K})\bigg{]}^{2}}{2\bigg{[} P^{(n)}_{(\lambda)}(\mathbf{K}) \bigg{]}^{2}}  \, .
\ea
Comparing the above optimal estimator, Eq. (\ref{optimales}), with the minimum estimator, Eq. (\ref{minimumes}), we  see that the correlation of the density fields with the fossil tensor modes appears as a four point function in terms of the density fields. Now, according to the above estimators, a signal is detected if the null hypothesis estimators defined above is found to depart at $>3 \sigma$ from the null hypothesis.

Now we evaluate the \textit{smallest} amplitude $A_{h}$ which can be detected within a given survey. In order to do this, we  assume that the fiducial power spectrum is nearly scale-invariant, say $P_{h}(\mathbf{K}) = A_h K^{n_h -3}$  with $n_h \simeq 0$. We also use our results for the squeezed limit of the fossil-density-density field bispectrum, Eq. (\ref{powerasymmetrythreepoint}). We then take the continuous limit of the summation in Eq. (\ref{noisepower}), i.e. by using $\sum_{\mathbf{k}} \longrightarrow V \int d^3k / (2\pi)^3$. We also  assume that for  $k<k_{max}$ we have $\left(\frac{P}{P^{tot}} \right) \sim 1$, where $k_{max}$ is the largest wave number for which the measured power spectrum would have a large signal to noise, and for any $k>k_{max}$ this ratio is equal to zero, i.e. we are dominated by the noise. Finally, we  also have to take into account  the direction-dependence of the tensor field. We consider the most generic case, given in  Appendix A,  and investigate the observability of a signal for this generic case. Then at the end of the day in order to calculate $\sigma$ we need to perform averaging over different directions of  GW. With this discussion 
in mind $P^{(n)}_{(\lambda)}(\mathbf{K})$ would be,
\ba
\label{Pnplus}
(P^{(n)}_{+}(\mathbf{K}))^{-1} &=& \left(\frac{g_{*}^2 N_{e}^2}{40 \pi^2} k_{max}^3\right) \left(
7 \cos^4{\theta}\cos^4{\phi} + 14 \cos^4{\theta}\cos^2{\phi} - 13 \cos^2{\theta}\cos^2{\phi} + 7 \cos^4{\theta} -15 \cos^2{\theta} +8
\right) \\
(P^{(n)}_{\times}(\mathbf{K}))^{-1} &=& \left(\frac{g_{*}^2 N_{e}^2}{40 \pi^2} k_{max}^3\right) \left(
 -28 \cos^4{\theta}\cos^2{\phi} + 29 \cos^2{\theta}\cos^2{\phi} - \cos^2{\theta} + 1
\right)  \, .
\ea
We can then plug back the final expression for the noise power-spectrum into Eq. (\ref{optimalvar}) in order for  the tensor amplitude to be detectable at the $\geq 3 \sigma$, 
\ba
3\sigma_{h} = \left(\frac{113\pi}{g_{*}^2 N_{e}^2}\right) \sqrt{\pi} \left(\frac{k_{max}}{k_{min}}\right)^{-3} \, .
\ea
So in the presence of the tensor mode with the amplitude $A_{t} = 3\times 10^{-10}$ (corresponding to tensor to scalar power spectra 
ratio $\sim 0.1$ )  in the matter two point function, 
$k_{max}/k_{min}$ would be 
\ba
\label{sensitivity}
k_{max}/k_{min} &=& 12800 (g_{*} N_{e})^{(-2/3)} \, .
\ea
For $g_{*}N_{e} \sim 2 $ and $g_{*}N_{e} \sim 8$  we obtain respectively $k_{max}/k_{min} \sim 8000$ and
$k_{max}/k_{min} \sim 3876$.
These predictions seem to be detectable  in the galaxy surveys like EUCLID and/or the 21cm surveys.

\subsection{Estimator for the Preferred Direction}

So far we had assumed that we  know the direction of anisotropy in the sky from another experiment. However, it might be that there is no  knowledge about the direction of anisotropy in the sky. So in this case it is worth to determine the shape of the estimator that we tried to calculate. For this purpose, we start from Eq. (\ref{minimumes}) and  construct the following estimator, 
\ba
h_{(\lambda)}^{lm}(\mathbf{K}) &=& \int d \hat{x}^2 \widetilde{h_{(\lambda)}(\mathbf{K})} Y_{lm}^{*}(\mathbf{\hat{x}}) \nonumber\\
& \sim & \int d \hat{x}^2 P^{n}_{(\lambda)}(\mathbf{K}, \mathbf{\hat{x}}) \sum_{\mathbf{k}}\frac{\Delta^{ij}(\mathbf{\hat{x}},\mathbf{\hat{k}}, \mathbf{\hat{K}-\hat{k}}) e^{(\lambda)}_{ij}(\mathbf{K})}{2V} \, 
Y_{lm}^{*}(\mathbf{\hat{x}})
\ea
Since in this case the  direction of anisotropy  is not  known we have written explicitly the dependency of  $P^{(n)}_{\lambda}(\mathbf{K}, \mathbf{\hat{x}})$ on the preferred direction, $ \mathbf{\hat{x}} $.

In the following, we try to approximate the above integral and figure out which values of  $l$ would be non-zero. For this purpose, it is convenient to neglect the dependency of $P^{(n)}_{(\lambda)}(\mathbf{K}, \mathbf{\hat{x}})$ on the preferred direction. So we are left with the following integral, 
\ba
\label{Shape}
h_{p}^{lm}(\mathbf{K}) \sim  \int d \hat{x}^2  \left(\mathbf{\hat{x}^i} \mathbf{\hat{x}^j} \right) Y_{lm}^{*}(\mathbf{\hat{x}}) \, 
\ea
where we have neglected the denominator as well as the summation over $k$ and the tensor. 
It is straightforward to show that the above estimator looks like a quadrupole. In order to show this, one should rewrite  $\hat{x}^i$ in terms of  $Y_{1m}$s. Then according to the additional momentum theorem, we see that the final value of $l$ can be only 0, 1 or 2. However, since the integral of the three spherical harmonics is proportional to $\left( \begin{array}{ccc}
l & 1 & 1\\
0 & 0 & 0
\end{array} \right)$, we  see that for $l=1$ this Wigner three-j symbol is zero. So we are  left with either $l=0$ or $l=2$. Among these two different cases, the first one is completely constant so we can imagine that it  contributes  to trace of the $e_{ij}$ which is zero. So at the end we are  left only with $l=2$  which proves that the result indeed looks like a quadrupole, 
\ba
\label{Shape}
h_{p}^{lm}(\mathbf{K}) \propto \left( \begin{array}{ccc}
	2 & 1 & 1\\
	0 & 0 & 0
\end{array} \right) \, .
\ea

\section{Conclusion and discussion}

While the  searches for the primordial GW in CMB 
are ongoing, it is worth to consider the large scale structure surveys for shedding light on the imprints of primordial GW. The goal of these surveys is to study the two point correlations of the density fluctuations. According to the null hypothesis, there is no off-diagonal correlation between different Fourier modes of the density fluctuations. However, in the presence of a long wavelength tensor mode different Fourier modes  couple to each other. As a result this off-diagonal correlation function can be thought as a way to probe the primordial GW. On the other hand, since the key point in these surveys is the power spectrum of the density perturbations, there should be a hierarchy between the scale of the tensor mode, assumed to  have a long wavelength, as compared with the typical scale of the density perturbations, supposed to have small wavelength. In another word one should consider the squeezed limit of the Tensor-Scalar-Scalar bispectrum. However, the tensor mode wavelength can be larger, comparable or smaller than the Hubble scale at any time. It has been proven that while in the first two cases the amplitude of GW  is still nearly constant, it is not the case for the last case, i.e. when the tensor mode re-enters  the horizon. In this case, it starts oscillating and decay but even in this case its effect survives even after the decay of the tensor mode \cite{Dai:2013kra}. That is why this effect is called  the fossil effect. Moreover, the shape of this off-diagonal correlation function looks like  a quadrupole anisotropy. 

There are three different contributions in the above TSS coming from the inflation, the non-linear mode coupling during the radiation and the matter domination era and finally the projection effect. It has been shown previously that for single field slow-roll inflationary models, there is a cancellation between the late time projection effect and both of the primordial inflationary contribution as well the non-linear mode coupling in the super horizon and sub horizon limit respectively. Therefore, the final correlation function is small in these models. However, the above cancellation does not exist for the models beyond the SFSR inflation and thus the quadrupole anisotropy can be large enough in these models which can be also thought as a way to distinguish between the single field and multiple field models. 

On the other hand, in models of anisotropic inflation in which there is a preferred direction in the early Universe and the three dimensional rotational symmetry has been reduced to a planar two dimensional symmetry,
the primordial power spectrum has quadrupole anisotropy  in the absence of  any tensor mode. Therefore,  it is worth to consider these models in detail and figure out a way to distinguish between the predictions of these models and models in which the rotational symmetry is intact.

In this paper, we  have calculated the quadrupole correction in the power spectrum in models of anisotropic inflation constructed from the dynamics of a $U(1)$ gauge field  and have shown that this quadrupole is completely aligned along the preferred direction. We then considered the three point TSS correlation function and determined the shape of the quadrupole anisotropy for this case as well. As we have shown, in this case the quadrupole anisotropy is partially aligned along the preferred direction. This means that the models of anisotropic inflation 
have a unique and distinguishable signature in large scale surveys compared to models of inflation based on scalar fields.

Finally, we considered the observational consequences of this quadrupole anisotropy and  used it as a tool to estimate the contribution of 
 the primordial GW
 as well as to find the preferred direction. We have presented the optimal estimator for the amplitude of the gravitational waves and have
 shown explicitly that there is a window in the parameter space where there is a  $3 \sigma$ detection limit on the amplitude of 
 GW where the signal is detectable in the galaxy surveys like Euclid and 21-cm. 

\section*{Acknowledgment}
We are very grateful to Marc Kamionkowski for his collaboration during the course of this work as well as for his numerous insightful comments and suggestions.   R.E. is grateful to Johns Hopkins University for the hospitality during the progress of this work. The work of R.E. is supported in part by the Balzan foundation via the New College Oxford-Johns Hopkins Centre for Cosmological Studies. 
\appendix

\section{The most generic tensor polarization}
\label{generic tensor}

In this appendix, we  present the most general form of  polarization for the tensor modes. Given a wave number as $\bfk = k(\cos{\theta} \sin{\varphi}, \sin{\theta}\sin{\varphi}, \cos{\varphi})$,
the most general tensor polarizations $e^{+}_{ij}(\mathbf{k})$ and $e^{\times}_{ij}(\mathbf{k})$ are given by
\begin{align}
\label{polarization2}
e^{+}_{ij}(\mathbf{k}) =&\frac{1}{\sqrt{2}} \left( \begin{array}{ccc}
1 - \cos^2{\theta} - \cos^2{\theta} \cos^2{\varphi}  & -\sin{\theta}\cos{\theta}(1 + \cos^2{\varphi}) & \sin{\varphi}\cos{\varphi}\cos{\theta} \\
-\sin{\theta}\cos{\theta}(1 + \cos^2{\varphi}) & \cos^2{\theta}\cos^2{\varphi} + \cos^2{\theta} - \cos^2{\varphi} & \sin{\theta} \sin{\varphi}\cos{\varphi} \\
\sin{\varphi}\cos{\varphi}\cos{\theta} & \sin{\theta} \sin{\varphi}\cos{\varphi} & -\sin^2{\varphi} \\
\end{array} \right) \\
e^{\times}_{ij}(\mathbf{k}) =& \frac{1}{\sqrt{2}}\left( \begin{array}{ccc}
2\sin{\theta}\cos{\theta}\cos{\varphi} & -\cos{\varphi}(2\cos^2{\theta}-1) & -\sin{\varphi}\sin{\theta} \\
-\cos{\varphi}(2\cos^2{\theta}-1) & -2\sin{\theta} \cos{\theta} \cos{\varphi} & \sin{\varphi}\cos{\theta} \\
-\sin{\varphi}\sin{\theta} & \sin{\varphi}\cos{\theta} & 0 \\
\end{array} \right) ~.
\end{align}
However, as we mentioned in the text, still we have the freedom to assume the planar tensor modes for the whole of our discussions in this work, so we can simply put $\varphi = \frac{\pi}{2}$. With this choice the polarizations are given in 
Eqs. (\ref{polarization2+}) and (\ref{polarization2times}).

\section{Interaction Lagrangian in the real space}
\label{int-lagrang}

In this Appendix, we present the third order interaction Lagrangian in the real space. Our starting point is the matter Lagrangian,
\ba
\label{interaction Lagrangian1}
L_{int} = -\frac{a^4}{4}f(\phi)^2 F_{\mu \nu} F^{\mu \nu} \, .
\ea
The goal is to calculate the third order interaction Lagrangian which can be obtained by expanding the above action up to the third order around a given background. Since the resulting Lagrangian  contains a lot of terms, and some of them are irrelevant to our discussion here, we restrict our attention only on to the following five sets of interactions, i.e. $L^{SST}$, $L^{SSV}$, $L^{SVT}$, $L^{SVV}$ and $L^{VVT}$. In addition, as we will show in the next Appendix, since the magnetic field is exponentially suppressed compared to the electric field, in the following we only consider the time derivatives and do not write down the spatial derivatives.
\ba
\label{detail-LSST}
L^{SST} &=& - 24 R M_{P}^2
\left(\frac{a^2}{\eta^2}\right)  \zeta(x) \zeta(x) h_{xx}(x)\\
\label{detail-LSSV}
L^{SSV} &=& -8 \sqrt{6 R} M_{P} \left(\frac{a f}{\eta}\right) \zeta(x) \zeta(x) \delta A'_{1} (x) \\
\label{detail-LSVT}
L^{SVT} &=& 4 \sqrt{6 R} M_{P} \left(\frac{a f}{\eta}\right) \zeta(x) \left(\delta A'_1(x) h_{xx}(x) + \delta A'_2(x) h_{xy}(x) + \delta A'_3(x) h_{xz}(x) \right) \\
\label{detail-SVV}
L^{SVV} &=& 2 f^2 \zeta(x)
\left( (\delta A'_{1}(x) )^2 + (\delta A'_2(x))^2 + (\delta A'_3(x))^2 \right) \\
\label{detailVVT}
L^{VVT} &=& -\frac{1}{2} f^2 \bigg{(} h_{xx}(x)(\delta A'_{1}(x) )^2 + 2 h_{xy}(x) \delta A'_{1}(x)\delta A'_{2}(x) + 2 h_{xz}(x) \delta A'_{1}(x) \delta A'_{3}(x) + h_{yy}(x) (\delta A'_{2}(x))^2 \nonumber\\
&&+  2 h_{yz}(x) \delta A'_{2}(x) \delta A'_{3}(x) + h_{zz}(x)(\delta A'_{3}(x) )^2 
\bigg{)} \, ,
\ea
where we have used the following useful expressions,
\ba
\left(\frac{\partial f^2}{\partial \phi}\right)\delta \phi &=& 4 f^2 \zeta \\
\frac{\partial^2 f^2}{\partial \phi^2} \delta \phi^2 &=& 16 f^2 \zeta^2 \\
fA'_{x} &=& M_{P}\sqrt{6 R} (-\eta)^{-1} a \\
\phi &=& M_{P}\sqrt{\frac{2}{\epsilon_{H}}} \\
f&=& \left(\frac{\eta^2}{\eta_{e}^2}\right) \, .
\ea
The remaining part is  to go to the Fourier space which is standard and we do not present it here.

\section{Interaction Lagrangian in the Fourier space}
\label{Fourierspacegeneric}

In the following, we  present the second and third order actions in the Fourier space for the most general choice of the wavenumber as well as the vector and the tensor polarizations.
\subsection{Second ordered action}
Let us start with the second order action.

\ba
\label{LSVgeneric}
L_{SV} &=& -4 M_{P}\sqrt{6 R}\int d^3k \left(\frac{a f}{\eta}\right) \zeta_{\mathbf{
	-k}} \left(\sin{\theta_{k}} A'_{\mathbf{k}} + \cos{\theta_{k}}\cos{\varphi_{k}} B'_{\mathbf{k}} \right) \\
\label{LSTgeneric}
L_{ST} &=& - 6 \sqrt{2} R M_{P}^2 \int d^3k \left(\frac{a^2}{\eta^2}\right) \sin^2{\theta_{k}} \zeta_{\mathbf{-k}} h_{+\mathbf{k}}\\
\label{LVTgeneric}
L_{VT} &=& M_{P} \sqrt{3 R}
\int d^3k \left( \frac{af}{\eta}\right) \bigg{(}\sin{\theta_{k}}A'_{\mathbf{-k}} {h}_{+\mathbf{k}} + \sin{\theta_{k}}B'_{\mathbf{-k}} {h}_{\times \mathbf{k}} \bigg{)}
\ea
\subsection{Third ordered action}
Here we present the full third order interaction Lagrangian.
\ba
\label{SSTgeneric}
L^{SST} &=& -12 \sqrt{2} R M_{P}^2
\left(\frac{a^2}{\eta^2}\right) \int d^3p \int d^3q
\sin^2{\theta_{q}} \zeta_{(\mathbf{-p-q})} \zeta_{\mathbf{p}} h_{+ \mathbf{q}}\\
\label{SSVgeneric}
L^{SSV} &=& -8 \sqrt{6 R} M_{P} \left(\frac{a f}{\eta}\right) \int d^3p \int d^3q
\zeta_{(\mathbf{-p-q})} \zeta_\mathbf{p}
\bigg{(}
\sin{\theta_{q}} A'_{\mathbf{q}} + \cos{\theta_{q}} \cos{\varphi_{q}} B'_{\mathbf{q}}
 \bigg{)}
 \\
\label{SVTgeneric}
L^{SVT} &=& 2 \sqrt{12 R} M_{P} \left(\frac{a f}{\eta}\right) \int d^3p \int d^3q \zeta_{(\mathbf{-p-q})}
\bigg{(} \sin{\theta_{p}} \cos{(\theta_{p} - \theta_{q})} h_{+\mathbf{p}} A'_{\mathbf{q}} \nonumber\\
&& +\sin{\theta_{p}} \cos{\varphi_{q}}  \sin{(\theta_{p} - \theta_{q})} h_{+\mathbf{p}} B'_{\mathbf{q}}  + \sin{\theta_{p}} \sin{\varphi_{q}}
h_{\times \mathbf{p}} B'_{\mathbf{q}}
\bigg{)}
\\
\label{SVVgeneric}
L^{SVV} &=& 2 f^2 \int d^3p \int d^3q \zeta_{(\mathbf{-p-q})}
\bigg{(}
\cos{(\theta_{p} - \theta_{q})} A'_{\mathbf{p}} A'_{\mathbf{q}} - 2 \sin{(\theta_{p} - \theta_{q})} \cos{\varphi_{p}} A'_{\mathbf{q}} B'_{\mathbf{p}} + \bigg{(} \cos{(\theta_{p} - \theta_{q})} \times  \nonumber\\
&& \times \cos{\varphi_{p}} \cos{\varphi_{q}} + \sin{\varphi_{p}} \sin{\varphi_{q}} \bigg{)} B'_{\mathbf{q}} B'_{\mathbf{p}} \bigg{)}
\ea
\ba
\label{VVTgeneric}
L^{VVT} &=& - \frac{\sqrt{2}}{4} f^2 \int d^3p \int d^3q
\bigg{(}\bigg{(}
- \sin^2{\theta_{(p+q)}} \cos{(\theta_{p} + \theta_{q})}
+\sin{\theta_{(p+q)}} \cos{\theta_{(p+q)}} \sin{(\theta_{p} + \theta_{q})} + \cos{\theta_{p}} \cos{\theta_{q}}
\bigg{)}\nonumber\\
&& \times  h_{+(\mathbf{-p-q})} A'_{\mathbf{p}} A'_{\mathbf{q}} 
+ \bigg{(} \sin{\theta_{p}} \cos{\theta_{q}} \cos{\varphi_{q}}
- \cos^2{\theta_{(p+q)}} \cos{\varphi_{q}} \sin{(\theta_{p} + \theta_{q})}+ \sin{(2\theta_{(p+q)})} \sin{\theta_{p}} \sin{\theta_{q}} \cos{\varphi_{q}}\bigg{)} \nonumber\\
&&\times h_{+ (\mathbf{-p-q})} A'_{\mathbf{p}} B'_{\mathbf{q}} -\bigg{(} \cos^2{\theta_{(p+q)}} \cos{\varphi_{p}} \sin{(\theta_{p} + \theta_{q})}- \cos{\theta_{p}} \cos{\varphi_{p}} \sin{\theta_{q}} 
-\sin{(2\theta_{(p+q)})}\cos{\theta_{p}} \cos{\varphi_{p}} \cos{\theta_{q}}\bigg{)} \nonumber\\
&&\times h_{+ (\mathbf{-p-q})} A'_{\mathbf{q}} B'_{\mathbf{p}}+
\bigg{(} 2 \sin{\varphi_{q}} \cos{(\theta_{p+q} - \theta_{p})}\bigg{)} h_{\times (\mathbf{-p-q})} A'_{\mathbf{p}} B'_{\mathbf{q}} 
+ \bigg{(}
\cos{\theta_{p}} \cos{\varphi_{p}} \cos{\theta_{q}} \cos{\varphi_{q}} - \sin^2{\varphi_{p}} \sin^2{\varphi_{q}} \nonumber\\
&&
- \cos^2{\theta_{(p+q)}} \cos{\varphi_{p}} \cos{\varphi_{q}} \cos{(\theta_{p} + \theta_{q})} 
 -\sin{(2\theta_{(p+q)})} \cos{\theta_{p}} \cos{\varphi_{p}} \sin{\theta_{q}} \cos{\varphi_{q}} 
\bigg{)} h_{+ (\mathbf{-p-q})} B'_{\mathbf{q}} B'_{\mathbf{p}} \nonumber\\
&&+ 2 \cos{\varphi_{p}} \sin{\varphi_{q}} \sin{(\theta_{(p+q)} - \theta_{p})} h_{\times (\mathbf{-p-q})} B'_{\mathbf{q}} B'_{\mathbf{p}} \bigg{)} \, .
\ea

\section{The in-in analysis for the full planar case}
\label{in-in}
In the following, we present the in-in formalism for the simplest case, which is the planar case.
\subsection{In-In integrals for the leading contributions in the full planar case}

In the following, by using the in-in formalism, we obtain the tensor-scalar-scalar bispectrum for each of the related diagrams, as given in the Fig. \ref{Feynmann}.\\
\begin{itemize}
	\item \textbf{SST}: In this case, there is only one possibility, taking into account that we are using the first order in-in formalism here. Therefore, the TSS bispectrum would be,
	\ba
	\left\langle h_{+\mathbf{k_1}} \zeta_{\mathbf{k_2}} \zeta_{\mathbf{k_3}}(\eta_e) \right\rangle &=&
	-24 \sqrt{2} R \left(\frac{M_P}{H}\right)^2 \sin^2{\theta_{k_1}} \int_{\eta_0}^{\eta_e} \frac{d\eta}{\eta^4} \nonumber\\
	&&\bigg{(}Im{\left(\zeta(k_2,\eta)\zeta^{*}(k_2,\eta_e) \zeta(k_3,\eta)\zeta^{*}(k_3,\eta_e) h_{+}(k_1, \eta)h^{*}_{+}(k_1, \eta_e)\right)} + (k_2 \Leftrightarrow k_3) \bigg{)} \nonumber\\
	&=& -
	\left(\frac{\sqrt{2}R N_e}{\epsilon_H}\right)
	\left(\sin^2{\theta_{k_1}} \left(\frac{k_1^3 + k_2^3 + k_3^3}{k_3^3}\right) P_{h_+}(\mathbf{k_1}) P_{\zeta}(\mathbf{k_2}) + (\mathbf{k_2} \Leftrightarrow \mathbf{k_3})\right) \, ,
	\ea
	where $\eta_0$ refers to an arbitrary time while $\eta_e$ denotes the end of inflation time.
	
	\item \textbf{SSV}: Here we  need to go to the second order in the in-in formalism. There are two different places for the interaction Hamiltonians to sit. They can either choose $\eta_1$ or $\eta_2$. However, a detailed analysis shows that the final results would be the same for both of them so we can just calculate one of them and keep in mind that the final result needs an extra factor 2, as has been written in the main text.
	\ba
	\left\langle h_{+\mathbf{k_1}} \zeta_{\mathbf{k_2}} \zeta_{\mathbf{k_3}} (\eta_e)\right\rangle &=&
	-96\sqrt{2} R \left(\frac{M_P}{H \eta_e^2}\right)^2 \sin^2{\theta_{k_1}} \int_{\eta_0}^{\eta_e} d\eta_1 \int_{\eta_0}^{\eta_1} d\eta_2
	\bigg{(}Im{\left(h_{+}(k_1, \eta_1)h^{*}_{+}(k_1, \eta_e)\right)} \nonumber\\
	&&
	 Im{\left(\zeta(k_2,\eta_2)\zeta^{*}(k_2,\eta_e) \zeta(k_3,\eta_2)\zeta^{*}(k_3,\eta_e) A'(k_1, \eta_2)A^{'*}(k_1, \eta_1)\right)}+ (k_2 \Leftrightarrow k_3)\bigg{)}\nonumber\\
	&=& -
	\left(\frac{\sqrt{2} R }{\epsilon_H}\right)
	\left(3 N_e^2 - 2 N_e \right)
    \sin^2{\theta_{k_1}}
	\bigg{(}\left(\frac{k_2^3 + k_3^3}{k_3^3}\right)
	P_{\zeta}(\mathbf{k_2})P_{h_+}(\mathbf{k_1}) \nonumber\\
	&& + (\mathbf{k_2} \Leftrightarrow \mathbf{k_3})\bigg{)} \, .
	\ea
	\item \textbf{SVT}: In this case, we need to consider the second order in-in formalism as well. Again there are two different locations for each of the Hamiltonian to sit. However, a detailed analysis  shows that they will be the same. Therefore in the following, we consider only one of them while keeping in mind that the final result would be twice, as is given in the main text
	\ba
	\left\langle h_{+\mathbf{k_1}} \zeta_{\mathbf{k_2}} \zeta_{\mathbf{k_3}} (\eta_e)\right\rangle &=&
	-192 \sqrt{2} R \left(\frac{M_P}{H \eta_e^2}\right)^2 \int_{\eta_0}^{\eta_e} d\eta_1 \int_{\eta_0}^{\eta_1} d\eta_2
	\sin{\theta_{k_1}} \bigg{(}\sin{\theta_{k_2}} \cos{(\theta_{k_1}- \theta_{k_2})}
	Im{\left(\zeta(k_2, \eta_1)\zeta^{*}(k_2, \eta_e)\right)} \nonumber\\
	&&
	Im{\left(\zeta(k_3, \eta_2)\zeta^{*}(k_3, \eta_e)
	h_{+}(k_1, \eta_2)h^{*}_{+}(k_1, \eta_e) A'(k_2, \eta_2)A^{'*}(k_2, \eta_1) \right)}
	+ (k_2 \Leftrightarrow k_3) \bigg{)}
	\nonumber\\
	&=& -
	\left(\frac{2 \sqrt{2} R}{\epsilon_H}\right)
	\left(3N_e^2 - 2 N_e \right)
	\bigg{(} \sin{\theta_{k_1}}\sin{\theta_{k_2}} \cos{(\theta_{k_1}- \theta_{k_2} )}\left(\frac{k_1^3 + k_3^3}{k_3^3}\right)
	\nonumber\\
	&& P_{\zeta}(\mathbf{k_2})P_{h_+}(\mathbf{k_1}) + (\mathbf{k_2} \Leftrightarrow \mathbf{k_3})\bigg{)} \, .
	\ea
	\item \textbf{VVT}: Here, one  needs to expand the in-in formalism up to the third order in the perturbations. In this case, there are three different positions for each of the interaction Hamiltonians. However, as we have already mentioned above, these different locations lead to the same results, so we can only consider one of them here and remember that the final result needs an additional factor of three, as is given in the main text
	\ba
	\left\langle h_{+\mathbf{k_1}} \zeta_{\mathbf{k_2}} \zeta_{\mathbf{k_3}} (\eta_e)\right\rangle &=&
	- 384 \sqrt{2} R \left(\frac{M_P}{H \eta_e^2}\right)^2 \int_{\eta_0}^{\eta_e} d\eta_1 \int_{\eta_0}^{\eta_1} d\eta_2 \int_{\eta_0}^{\eta_2} d\eta_3 \left(\frac{\eta_3}{\eta_e}\right)^4
	\bigg{(}
	\sin{\theta_{k_2}}\sin{\theta_{k_3}} \cos{(\theta_{k_1}- \theta_{k_2} )}\nonumber\\
	&& \cos{(\theta_{k_1}- \theta_{k_3} )}
	Im{\left(\zeta(k_2, \eta_1)\zeta^{*}(k_2, \eta_e)\right)} Im{\left(\zeta(k_3, \eta_2)\zeta^{*}(k_3, \eta_e)\right)} \nonumber\\
	&&
	Im{\left( h_{+}(k_1, \eta_3)h^{*}_{+}(k_1, \eta_e) A'(k_3, \eta_3)A^{'*}(k_3, \eta_2) A'(k_2, \eta_3)A^{'*}(k_2, \eta_1) \right)}
	+ (k_2 \Leftrightarrow k_3)
	\bigg{)}
	\nonumber\\
	&=& -
	\left(\frac{6 \sqrt{2} R }{\epsilon_H}\right)
	\left( N_e^3 - N_e^2  + \frac{1}{3} N_e\right)
	\bigg{(} \sin{\theta_{k_2}}\sin{\theta_{k_3}} \cos{(\theta_{k_1}- \theta_{k_2} )} \cos{(\theta_{k_1}- \theta_{k_3} )}
	\nonumber\\ && \left(\frac{k_1^3}{k_3^3}\right)P_{\zeta}(\mathbf{k_2})P_{h_+}(\mathbf{k_1}) + (\mathbf{k_2} \Leftrightarrow \mathbf{k_3})\bigg{)} \, .
	\ea
	\item \textbf{SVV}: Like in the previous case, one needs to consider the third order in-in formalism for this case. There are six different positions for the interaction Hamiltonians in this case. However, they would be the same and as in the above cases, it is enough to only consider one of them and keep in mind than the final result  needs an additional factor 6, as it is given in the main text
	\ba
	\label{SVVin-in}
	\left\langle h_{+\mathbf{k_1}} \zeta_{\mathbf{k_2}} \zeta_{\mathbf{k_3}} (\eta_e)\right\rangle &=&
	- 384 \sqrt{2} R \left(\frac{M_P}{H \eta_e^2}\right)^2 \int_{\eta_0}^{\eta_e} d\eta_1 \int_{\eta_0}^{\eta_1} d\eta_2 \int_{\eta_0}^{\eta_2} d\eta_3 \left(\frac{\eta_3}{\eta_e}\right)^4
	\bigg{(}
	\sin{\theta_{k_1}}\sin{\theta_{k_2}} \cos{(\theta_{k_1}- \theta_{k_2} )}\nonumber\\
	&&
	Im{\left(h_{+}(k_1, \eta_1)h_{+}^{*}(k_1, \eta_e)\right)} Im{\left(\zeta(k_2, \eta_2)\zeta^{*}(k_2, \eta_e)\right)} \nonumber\\
	&&
	Im{\left( \zeta(k_3, \eta_3)\zeta^{*}(k_3, \eta_e) A'(k_1, \eta_3)A^{'*}(k_1, \eta_1) A'(k_2, \eta_3)A^{'*}(k_2, \eta_2) \right)}
	+ (k_2 \Leftrightarrow k_3)
	\bigg{)}
	\nonumber\\
	&=& -
	\left( 	\frac{6 \sqrt{2} R }{\epsilon_H} \right)
	\left( N_e^3 - N_e^2  + \frac{1}{3} N_e \right)
	\bigg{(} \sin{\theta_{k_1}}\sin{\theta_{k_2}} \cos{(\theta_{k_1}- \theta_{k_2} )}
	\nonumber\\ &&
	P_{\zeta}(\mathbf{k_2})P_{h_+}(\mathbf{k_1}) + (\mathbf{k_2} \Leftrightarrow \mathbf{k_3})\bigg{)}
	\ea
\end{itemize}
\subsection{Proving the suppression of the terms without time derivative of the gauge field}
\label{magnetic field suppression}

As it has been argued throughout the text, we do not consider terms which do not contain the time derivative of the gauge field. One intuitive reason might be due to the fact that magnetic field decays outside the horizon. So one can neglect them and only consider the electric field, as we did it for the whole of the analysis. However, in order to be more precise, it is worth to do a sample analysis for this case as well to have a better feeling about the order of magnitude of these terms. So in the following, we consider one example of the magnetic field, for example
let us consider the following interaction in the SVV sector
\ba
L^{SVV}_{magnetic field} = 2 f^2 \int d^3p \int d^3q pq \zeta_{(\mathbf{-p-q})} A_{\mathbf{p}} A_{\mathbf{q}}
\ea
Using the in-in formalism, we  have the following expression for the tensor-scalar-scalar bispectrum,
\ba
\label{magnetic field}
\left\langle h_{+\mathbf{k_1}} \zeta_{\mathbf{k_2}} \zeta_{\mathbf{k_3}} (\eta_e)\right\rangle &=&
- \left(384 \sqrt{2} R\right) \left(\frac{M_P}{H \eta_e^2}\right)^2 \int_{\eta_0}^{\eta_e} d\eta_1 \int_{\eta_0}^{\eta_1} d\eta_2 \int_{\eta_0}^{\eta_2} d\eta_3 \left(\frac{\eta_1}{\eta_e}\right)^4
\bigg{(}
k_1 k_2 \sin{\theta_{k_1}}\sin{\theta_{k_2}}\nonumber\\
&&
Im{\left(\zeta(k_3, \eta_1)\zeta^{*}(k_3, \eta_e)\right)} Im{\left(\zeta(k_2, \eta_2)\zeta^{*}(k_2, \eta_e) A'(k_2, \eta_2)A^{*}(k_2, \eta_1) \right)} \nonumber\\
&& Im{\left( h_{+}(k_1, \eta_3)h_{+}^{*}(k_1, \eta_e) A'(k_1, \eta_3)A^{*}(k_1, \eta_1)\right)}
+ (k_2 \Leftrightarrow k_3) \bigg{)} \nonumber\\
&=& - \left(\frac{27 \sqrt{2} R}{100 \epsilon_H} \right)
\bigg{(} \sin{\theta_{k_1}}\sin{\theta_{k_2}} \left(k_1\eta_0\right) \left(k_2\eta_0\right) P_{\zeta}(\mathbf{k_2})P_{h_+}(\mathbf{k_1}) + (\mathbf{k_2} \Leftrightarrow \mathbf{k_3})\bigg{)} \, .
\ea
Comparing Eq. (\ref{magnetic field}) with Eq. (\ref{SVVin-in}), we see that the contribution from the magnetic field is indeed suppressed as compared to the electric field and it is safe to neglect it for the whole of our analysis.

\section{Detailed Analysis of the root-mean-square of the Quadrupole Asymmetry}
\label{quadrupole rms}

In this Appendix, we  calculate the root-mean-square of the quadrupole anisotropy within our anisotropic setup.

Starting from the definition of the root-mean-square of the Quadrupole asymmetry, 
\ba
\label{root-mean-square-def}
\overline{Q}^2 \equiv \bigg{\langle} \sum_{M=-2}^{+2} | Q_{2M}|^2 \bigg{\rangle} \, ,
\ea
we  need to calculate the following quantity, 
\ba
\label{rms first step}
\overline{Q}^2 = \int d^2 \hat{n} \int d^2 \hat{n'} \Delta^{ij}(\mathbf{\hat{x}}, \mathbf{\hat{n}}) \Delta^{ab}(\mathbf{\hat{x}}, \mathbf{\hat{n}'}) \sum_{M=-2}^{+2}  Y_{2M}(\mathbf{\hat{n}'}) Y^{*}_{2M}(\mathbf{\hat{n}}) \bigg{\langle} h_{ij}(\mathbf{K_{L}}) h_{ab}(\mathbf{K_{L}}) \bigg{\rangle} \, .
\ea
Now, in order to proceed, we can use the following formula for the GW \cite{Ruth Durrer}, 
\ba
\label{twopoint tensor}
\bigg{\langle} h_{ij}(\mathbf{K_{L}}) h_{ab}(\mathbf{K_{L}}) \bigg{\rangle} = P_{h}(\mathbf{K_{L}}) M_{ijab}(\mathbf{K_{L}}) \, ,
\ea
where $M_{ijab}(\mathbf{K_{L}})$ is defined in the following way,
\ba
\label{Mijab}
M_{ijab}(\mathbf{K_{L}}) \equiv \left( P_{ia} P_{jb} + P_{ib} P_{ja} - P_{ij} P_{ab} \right) \, ,
\ea
in which $P_{ij} = (\delta_{ij} - \hat{K}_{Li}\hat{K}_{Lj})$ is the projection tensor into the surface perpendicular to the GW wave number.

Plugging back the above definition for the projection tensor inside $M_{ijab}(K_{L})$, we would have,
\ba
\label{twopoint tensor final}
\bigg{\langle} h_{ij}(\mathbf{K_{L}}) h_{ab}(\mathbf{K_{L}}) \bigg{\rangle} &=& P_{h}(K_{L})\bigg{(}
\delta_{ia}\delta_{jb} + \delta_{ib}\delta_{ja} - \delta_{ij}\delta_{ab} + \delta_{ij} \hat{K}_{La}\hat{K}_{Lb} + \delta_{ab} \hat{K}_{Li}\hat{K}_{Lj} - \delta_{ia} \hat{K}_{Lj}\hat{K}_{Lb} \nonumber\\
&& - \delta_{jb} \hat{K}_{Li}\hat{K}_{La} - \delta_{ib} \hat{K}_{Lj}\hat{K}_{La} - \delta_{ja} \hat{K}_{Li}\hat{K}_{Lb} + \hat{K}_{Li}\hat{K}_{Lj}\hat{K}_{La}\hat{K}_{Lb}
 \bigg{)} \, .
\ea
We also recall the expression for the $\Delta_{ij}(\mathbf{\hat{x}} , \mathbf{\hat{n}})$, 
\ba
\Delta_{ij}(\mathbf{\hat{x}} , \mathbf{\hat{n}}) = 3g_{*}N_{e} \left( \hat{x}^{i} \hat{x}^{j} - (\mathbf{\hat{n}}\cdot \mathbf{\hat{x}}) \hat{x}^{i} \hat{n}^{j}\right) \, .
\ea

Plugging the above expressions back into the $\overline{Q}^2$, and using the following expression for the product of two $Y_{LM}$, 
\ba
\label{sumation rule}
\sum_{M=-L}^{L} Y_{LM}(\mathbf{\hat{n}}) Y_{LM}^{*}(\mathbf{\hat{n}'}) = \left(\frac{2L+1}{4\pi}\right) P_{L}(\mathbf{\hat{n}}\cdot \mathbf{\hat{n}'}) \, ,
\ea
 we would get,
\ba
\label{root-mean-squre third}
\overline{Q}^2 &=& \left(\frac{45}{4\pi}\right) \left(g_{*}N_{e}\right)^2 P_{h}(K_{L}) \int d^2 \hat{n} \int d^2 \hat{n}' P_{2}(\mathbf{\hat{n}}\cdot \mathbf{\hat{n}'}) \bigg{(} 1 -(\mathbf{\hat{n}}\cdot \mathbf{\hat{x})}^2 - (\mathbf{\hat{n}'}\cdot \mathbf{\hat{x}})^2 + (\mathbf{\hat{n}}\cdot \mathbf{\hat{x}}) (\mathbf{\hat{n}'}\cdot \mathbf{\hat{x}})(\mathbf{\hat{n}}\cdot \mathbf{\hat{n}'}) -2 (\mathbf{\hat{K}_{L}}\cdot \mathbf{\hat{x}})^2 \nonumber\\
&& + (\mathbf{\hat{n}'}\cdot \mathbf{\hat{x}}) (\mathbf{\hat{K}_{L}}\cdot \mathbf{\hat{x}}) (\mathbf{\hat{K}_{L}}\cdot \mathbf{\hat{n}'}) + (\mathbf{\hat{n}}\cdot \mathbf{\hat{x}}) (\mathbf{\hat{K}_{L}}\cdot \mathbf{\hat{x}}) (\mathbf{\hat{K}_{L}}\cdot \mathbf{\hat{n}}) - (\mathbf{\hat{n}}\cdot \mathbf{\hat{x}}) (\mathbf{\hat{n}'}\cdot \mathbf{\hat{x}}) (\mathbf{\hat{K}_{L}}\cdot \mathbf{\hat{n}}) (\mathbf{\hat{K}_{L}}\cdot \mathbf{\hat{n}'}) + (\mathbf{\hat{n}'}\cdot \mathbf{\hat{x}})^2 (\mathbf{\hat{K}_{L}}\cdot \mathbf{\hat{x}})^2 \nonumber\\
&& + (\mathbf{\hat{n}}\cdot \mathbf{\hat{x}})^2 (\mathbf{\hat{K}_{L}}\cdot \mathbf{\hat{x}})^2
-(\mathbf{\hat{n}}\cdot \mathbf{\hat{x}})(\mathbf{\hat{n}'}\cdot \mathbf{\hat{x}})(\mathbf{\hat{n}}\cdot \mathbf{\hat{n}'})(\mathbf{\hat{K}_{L}}\cdot \mathbf{\hat{x}})^2 + (\mathbf{\hat{K}_{L}}\cdot \mathbf{\hat{x}})^4 - (\mathbf{\hat{n}'}\cdot \mathbf{\hat{x}}) (\mathbf{\hat{K}_{L}}\cdot \mathbf{\hat{x}})^3 (\mathbf{\hat{K}_{L}}\cdot \mathbf{\hat{n}'}) \nonumber\\
&& - (\mathbf{\hat{n}}\cdot \mathbf{\hat{x}}) (\mathbf{\hat{K}_{L}}\cdot \mathbf{\hat{x}})^3 (\mathbf{\hat{K}_{L}}\cdot \mathbf{\hat{n}}) + (\mathbf{\hat{n}}\cdot \mathbf{\hat{x}})(\mathbf{\hat{n}'}\cdot \mathbf{\hat{x}}) (\mathbf{\hat{K}_{L}}\cdot \mathbf{\hat{x}})^2 (\mathbf{\hat{K}_{L}}\cdot \mathbf{\hat{n}}) (\mathbf{\hat{K}_{L}}\cdot \mathbf{\hat{n}'})
\bigg{)}  \, .
\ea
Now the remaining part is to calculate the above integrals. For this purpose, we  use the following formulas,
\ba
\label{useful expressions}
\int d^2 \hat{k} P_{l}(\mathbf{\hat{k}}\cdot \mathbf{\hat{n}}) P_{l'}(\mathbf{\hat{k}}\cdot \mathbf{\hat{n}'}) &=& \left(\frac{4\pi}{2l+1}\right)\delta_{ll'} P_{l}(\mathbf{\hat{n}}\cdot \mathbf{\hat{n}'}) \\
xP_{l}(x) &=& \left(\frac{l+1}{2l+1}\right)P_{l+1}(x) + \left(\frac{l}{2l+1}\right)P_{l-1}(x) 
\ea
as well as Eq. (\ref{sumation rule}), to rewrite $P_{l}(\mathbf{\hat{n}}\cdot \mathbf{\hat{n}'})$ in terms of the products of $Y_{lm}$s.

We find the final result for the root-mean-square of the quadrupole anisotropy as,
\ba
\label{root-mean-square finall}
\overline{Q}^2 &=& (8\pi) \left(g_{*}N_{e}\right)^2 P_{h}(K_{L}) \left(1 - (\mathbf{\hat{K_{L}}}\cdot \mathbf{\hat{x}})^2\right)\bigg{(} 1 - \left(40 \pi^2\right) \sum_{s,s',q,q' =-1}^{+1} \sum_{t=-2}^{+2} Y_{1s}^{*}(\mathbf{\hat{x}}) Y_{1s'}^{*}(\mathbf{\hat{x}}) Y_{1q}^{*}(\mathbf{\hat{K_{L}}}) Y_{1q'}^{*}(\mathbf{\hat{K_{L}}}) \nonumber\\
&&\left( \begin{array}{ccc}
	2 & 1 & 1\\
	0 & 0 & 0
\end{array} \right)^2
\left( \begin{array}{ccc}
	2 & 1 & 1\\
	t & q & s
\end{array} \right)
\left( \begin{array}{ccc}
	2 & 1 & 1\\
	-t & q' & s'
\end{array} \right)
\bigg{)}
\ea
As we see from Eq. (\ref{root-mean-square finall}), the final result is a function of  $\mathbf{K_{L}}$ as well as the preferred direction. So we may also take the average of this function with respect to either $\mathbf{K_{L}}$ or the preferred direction, $\mathbf{\hat{x}}$.




\section*{References}\vspace{-1cm}
\begin{thebibliography}{999}

\bibitem{Ade:2015lrj} 
  P.~A.~R.~Ade {\it et al.}  [Planck Collaboration],
  arXiv:1502.02114 [astro-ph.CO].

	
	\bibitem{Jeong:2012df} 
	D.~Jeong and M.~Kamionkowski,
	Phys.\ Rev.\ Lett.\  {\bf 108}, 251301 (2012)
	[arXiv:1203.0302 [astro-ph.CO]].
	
	\bibitem{Ade:2014xna} 
	P.~A.~R.~Ade {\it et al.}  [BICEP2 Collaboration],
	Phys.\ Rev.\ Lett.\  {\bf 112}, no. 24, 241101 (2014)
	[arXiv:1403.3985 [astro-ph.CO]].
	
	\bibitem{Ade:2015tva} 
	P.~A.~R.~Ade {\it et al.}  [BICEP2 and Planck Collaborations],
	Phys.\ Rev.\ Lett.\  {\bf 114}, no. 10, 101301 (2015)
	[arXiv:1502.00612 [astro-ph.CO]].
	
	\bibitem{Keisler:2015jua} 
	R.~Keisler, S.~Hoover, N.~Harrington, J.~W.~Henning, P.~A.~R.~Ade, K.~A.~Aird, J.~E.~Austermann and J.~A.~Beall {\it et al.},
	arXiv:1503.02315 [astro-ph.CO].
	
	
	\bibitem{Dodelson:2003bv} 
	S.~Dodelson, E.~Rozo and A.~Stebbins,
	Phys.\ Rev.\ Lett.\  {\bf 91}, 021301 (2003)
	[astro-ph/0301177].
	
	\bibitem{Schmidt:2012nw} 
	F.~Schmidt and D.~Jeong,
	Phys.\ Rev.\ D {\bf 86}, 083513 (2012)
	[arXiv:1205.1514 [astro-ph.CO]].
	\bibitem{Smith:2008pf} 
	T.~L.~Smith, M.~Kamionkowski and A.~Cooray,
	Phys.\ Rev.\ D {\bf 78}, 083525 (2008)
	[arXiv:0802.1530 [astro-ph]].
	
\bibitem{Pagano:2007st} 
L.~Pagano, A.~Cooray, A.~Melchiorri and M.~Kamionkowski,
JCAP {\bf 0804}, 009 (2008)
[arXiv:0707.2560 [astro-ph]].
	
\bibitem{Smith:2006nka} 
T.~L.~Smith, E.~Pierpaoli and M.~Kamionkowski,
Phys.\ Rev.\ Lett.\  {\bf 97}, 021301 (2006)
[astro-ph/0603144].
	
\bibitem{Pritchard:2004qp} 
J.~R.~Pritchard and M.~Kamionkowski,
Annals Phys.\  {\bf 318}, 2 (2005)
[astro-ph/0412581].

\bibitem{Kesden:2004qx} 
M.~Kesden, J.~Gair and M.~Kamionkowski,
Phys.\ Rev.\ D {\bf 71}, 044015 (2005)
[astro-ph/0411478].

\bibitem{Kesden:2002ku} 
M.~Kesden, A.~Cooray and M.~Kamionkowski,
Phys.\ Rev.\ Lett.\  {\bf 89}, 011304 (2002)
[astro-ph/0202434].

\bibitem{Kamionkowski:2000gb} 
M.~Kamionkowski and A.~H.~Jaffe,
Int.\ J.\ Mod.\ Phys.\ A {\bf 16S1A}, 116 (2001)
[astro-ph/0011329].

\bibitem{Caldwell:1998aa}
R.~R.~Caldwell, M.~Kamionkowski and L.~Wadley,
Phys.\ Rev.\ D {\bf 59} (1999) 027101
[astro-ph/9807319].
	
\bibitem{Kamionkowski:1997av} 
M.~Kamionkowski and A.~Kosowsky,
Phys.\ Rev.\ D {\bf 57}, 685 (1998)
[astro-ph/9705219].
	\bibitem{Dai:2012bc} 
	L.~Dai, M.~Kamionkowski and D.~Jeong,
	Phys.\ Rev.\ D {\bf 86}, 125013 (2012)
	[arXiv:1209.0761 [astro-ph.CO]].
	
	\bibitem{Chisari:2014xia} 
	N.~E.~Chisari, C.~Dvorkin and F.~Schmidt,
	Phys.\ Rev.\ D {\bf 90}, no. 4, 043527 (2014)
	[arXiv:1406.4871 [astro-ph.CO]].
	
\bibitem{Pen:2003yv} 
U.~L.~Pen,
New Astron.\  {\bf 9}, 417 (2004)
[astro-ph/0305387].

\bibitem{Masui:2010cz} 
K.~W.~Masui and U.~L.~Pen,
Phys.\ Rev.\ Lett.\  {\bf 105}, 161302 (2010)
[arXiv:1006.4181 [astro-ph.CO]].

\bibitem{Book:2011dz} 
L.~Book, M.~Kamionkowski and F.~Schmidt,
Phys.\ Rev.\ Lett.\  {\bf 108}, 211301 (2012)
[arXiv:1112.0567 [astro-ph.CO]].

\bibitem{Akhshik:2014gja} 
M.~Akhshik, R.~Emami, H.~Firouzjahi and Y.~Wang,
JCAP {\bf 1409}, 012 (2014)
[arXiv:1405.4179 [astro-ph.CO]].
	

	\bibitem{Dai:2013kra} 
	L.~Dai, D.~Jeong and M.~Kamionkowski,
	Phys.\ Rev.\ D {\bf 88}, no. 4, 043507 (2013)
	[arXiv:1306.3985 [astro-ph.CO]].
	
	\bibitem{Schmidt:2013gwa} 
	F.~Schmidt, E.~Pajer and M.~Zaldarriaga,
	Phys.\ Rev.\ D {\bf 89}, 083507 (2014)
	[arXiv:1312.5616 [astro-ph.CO]].
	
	\bibitem{Pajer:2013ana} 
	E.~Pajer, F.~Schmidt and M.~Zaldarriaga,
	Phys.\ Rev.\ D {\bf 88}, no. 8, 083502 (2013)
	[arXiv:1305.0824 [astro-ph.CO]].
	
	\bibitem{Maldacena:2002vr} 
	J.~M.~Maldacena,
	JHEP {\bf 0305}, 013 (2003)
	[astro-ph/0210603].
	
	
	\bibitem{Dimastrogiovanni:2014ina} 
	E.~Dimastrogiovanni, M.~Fasiello, D.~Jeong and M.~Kamionkowski,
	arXiv:1407.8204 [astro-ph.CO].
	
	\bibitem{Akhshik:2014bla} 
	M.~Akhshik,
	arXiv:1409.3004 [astro-ph.CO].
	
	\bibitem{Dimastrogiovanni:2015pla} 
	E.~Dimastrogiovanni, M.~Fasiello and M.~Kamionkowski,
	arXiv:1504.05993 [astro-ph.CO].
	
		
\bibitem{Watanabe:2009ct}
  M.~a.~Watanabe, S.~Kanno and J.~Soda,
  ``Inflationary Universe with Anisotropic Hair,''
  Phys.\ Rev.\ Lett.\  {\bf 102}, 191302 (2009)
  [arXiv:0902.2833 [hep-th]].
  
 \bibitem{Watanabe:2010fh}
  M.~a.~Watanabe, S.~Kanno and J.~Soda,
  ``The Nature of Primordial Fluctuations from Anisotropic Inflation,''
  Prog.\ Theor.\ Phys.\  {\bf 123}, 1041 (2010)
  [arXiv:1003.0056 [astro-ph.CO]].

 \bibitem{Soda1} 
J.~Ohashi, J.~Soda and S.~Tsujikawa,
``Observational signatures of anisotropic inflationary models,''
JCAP {\bf 1312}, 009 (2013)
[arXiv:1308.4488 [astro-ph.CO], arXiv:1308.4488].

J.~Ohashi, J.~Soda and S.~Tsujikawa,
``Anisotropic power-law k-inflation,''
Phys.\ Rev.\ D {\bf 88}, 103517 (2013)
[arXiv:1310.3053 [hep-th]].

J.~Ohashi, J.~Soda and S.~Tsujikawa,
``Anisotropic Non-Gaussianity from a Two-Form Field,''
Phys.\ Rev.\ D {\bf 87}, 083520 (2013)
[arXiv:1303.7340 [astro-ph.CO]].

S.~Kanno, J.~Soda, M.~-a.~Watanabe,
``Anisotropic Power-law Inflation,''
JCAP {\bf 1012}, 024 (2010).
[arXiv:1010.5307 [hep-th]].

K.~Murata, J.~Soda,
``Anisotropic Inflation with Non-Abelian Gauge Kinetic Function,''
JCAP {\bf 1106}, 037 (2011).
[arXiv:1103.6164 [hep-th]].

S.~Yokoyama and J.~Soda,
``Primordial statistical anisotropy generated at the end of inflation,''
JCAP {\bf 0808}, 005 (2008);

M.~-a.~Watanabe, S.~Kanno and J.~Soda,
``Imprints of Anisotropic Inflation on the Cosmic Microwave Background,''
Mon.\ Not.\ Roy.\ Astron.\ Soc.\  {\bf 412}, L83 (2011)
[arXiv:1011.3604 [astro-ph.CO]].

K.~Yamamoto, M.~-a.~Watanabe and J.~Soda,
``Inflation with Multi-Vector-Hair: The Fate of Anisotropy,''
Class.\ Quant.\ Grav.\  {\bf 29}, 145008 (2012)
[arXiv:1201.5309 [hep-th]].


\bibitem{Emami1}
R.~Emami, H.~Firouzjahi, S.~M.~Sadegh Movahed, M.~Zarei,
``Anisotropic Inflation from Charged Scalar Fields,''
JCAP {\bf 1102 } (2011)  005.
[arXiv:1010.5495 [astro-ph.CO]].

R.~Emami and H.~Firouzjahi,
``Issues on Generating Primordial Anisotropies at the End of Inflation,''
JCAP {\bf 1201}, 022 (2012)
[arXiv:1111.1919 [astro-ph.CO]].


S.~Baghram, M.~H.~Namjoo and H.~Firouzjahi,
``Large Scale Anisotropic Bias from Primordial non-Gaussianity,''
arXiv:1303.4368 [astro-ph.CO].

\bibitem{Emami:2013bk}
R.~Emami and H.~Firouzjahi,
``Curvature Perturbations in Anisotropic Inflation with Symmetry Breaking,''
JCAP {\bf 1310}, 041 (2013)
[arXiv:1301.1219 [hep-th]].

\bibitem{Abolhasani:2013zya}
A.~A.~Abolhasani, R.~Emami, J.~T.~Firouzjaee and H.~Firouzjahi,
``$\delta N$ formalism in anisotropic inflation and large anisotropic bispectrum and trispectrum,''
JCAP {\bf 1308}, 016 (2013)
[arXiv:1302.6986 [astro-ph.CO]].

\bibitem{Abolhasani:2013bpa} 
A.~A.~Abolhasani, R.~Emami and H.~Firouzjahi,
``Primordial Anisotropies in Gauged Hybrid Inflation,''
JCAP {\bf 1405}, 016 (2014)
[arXiv:1311.0493 [hep-th]].


\bibitem{Chen:2014eua} 
X.~Chen, R.~Emami, H.~Firouzjahi and Y.~Wang,
JCAP {\bf 1408}, 027 (2014)
[arXiv:1404.4083 [astro-ph.CO]].


\bibitem{Bartolo:2012sd}
  N.~Bartolo, S.~Matarrese, M.~Peloso and A.~Ricciardone,
  ``The anisotropic power spectrum and bispectrum in the $f(\phi) F^2$ mechanism,''
  Phys.\ Rev.\ D {\bf 87}, 023504 (2013)
  [arXiv:1210.3257 [astro-ph.CO]].
  
  \bibitem{Shiraishi:2013vja}
M.~Shiraishi, E.~Komatsu, M.~Peloso and N.~Barnaby,
``Signatures of anisotropic sources in the squeezed-limit bispectrum of the cosmic microwave background,''
JCAP {\bf 1305}, 002 (2013)
[arXiv:1302.3056 [astro-ph.CO]].
  
  \bibitem{Shiraishi:2013oqa}
M.~Shiraishi, E.~Komatsu and M.~Peloso,
``Signatures of anisotropic sources in the trispectrum of the cosmic microwave background,''
arXiv:1312.5221 [astro-ph.CO].


\bibitem{various}

  A.~E.~Gumrukcuoglu, B.~Himmetoglu, M.~Peloso,
  ``Scalar-Scalar, Scalar-Tensor, and Tensor-Tensor Correlators from Anisotropic Inflation,''
  Phys.\ Rev.\  {\bf D81}, 063528 (2010).
  [arXiv:1001.4088 [astro-ph.CO]].

T.~R.~Dulaney, M.~I.~Gresham,
``Primordial Power Spectra from Anisotropic Inflation,''
Phys.\ Rev.\  {\bf D81}, 103532 (2010).
[arXiv:1001.2301 [astro-ph.CO]].


H.~Funakoshi and K.~Yamamoto,
``Primordial bispectrum from inflation with background gauge fields,''
arXiv:1212.2615 [astro-ph.CO].

K.~Yamamoto,
``Primordial Fluctuations from Inflation with a Triad of Background Gauge Fields,''
Phys.\ Rev.\ D {\bf 85}, 123504 (2012)
[arXiv:1203.1071 [astro-ph.CO]].

  H.~Funakoshi and K.~Yamamoto,
  ``Primordial bispectrum from inflation with background gauge fields,''
  arXiv:1212.2615 [astro-ph.CO].
  
T.~Fujita and S.~Yokoyama,
``Higher order statistics of curvature perturbations in IFF model and its Planck constraints,''
JCAP {\bf 1309}, 009 (2013)
[arXiv:1306.2992 [astro-ph.CO]].


S.~R.~Ramazanov and G.~Rubtsov,
``Constraining anisotropic models of the early Universe with WMAP9 data,''
Phys.\ Rev.\ D {\bf 89}, 043517 (2014)
[arXiv:1311.3272 [astro-ph.CO]].

S.~Nurmi and M.~S.~Sloth,
``Constraints on Gauge Field Production during Inflation,''
arXiv:1312.4946 [astro-ph.CO].

R.~K.~Jain and M.~S.~Sloth,
``On the non-Gaussian correlation of the primordial curvature perturbation with vector fields,''
arXiv:1210.3461 [astro-ph.CO].

F.~R.~Urban,
``Pseudoscalar N-flation and axial coupling revisited,''
arXiv:1307.5215 [astro-ph.CO].

M.~Thorsrud, D.~F.~Mota and S.~Hervik,
``Cosmology of a Scalar Field Coupled to Matter and an Isotropy-Violating Maxwell Field,''
JHEP {\bf 1210}, 066 (2012)
[arXiv:1205.6261 [hep-th]].

S.~Bhowmick and S.~Mukherji,
``Anisotropic Power Law Inflation from Rolling Tachyons,''
Mod.\ Phys.\ Lett.\ A {\bf 27}, 1250009 (2012)
[arXiv:1105.4455 [hep-th]].

S.~Hervik, D.~F.~, M.~Thorsrud,
``Inflation with stable anisotropic hair: Is it cosmologically viable?,''
[arXiv:1109.3456 [gr-qc]].

C.~G.~Boehmer, D.~F.~Mota,
``CMB Anisotropies and Inflation from Non-Standard Spinors,''
Phys.\ Lett.\  {\bf B663}, 168-171 (2008).
[arXiv:0710.2003 [astro-ph]].

T.~S.~Koivisto, D.~F.~Mota,
``Vector Field Models of Inflation and Dark Energy,''
JCAP {\bf 0808}, 021 (2008).
[arXiv:0805.4229 [astro-ph]].

J.~P.~Beltran Almeida, Y.~Rodriguez and C.~A.~Valenzuela-Toledo,
``The Suyama-Yamaguchi consistency relation in the presence of
vector fields,''
Mod.\ Phys.\ Lett.\ A {\bf 28} (2013) 1350012
[arXiv:1112.6149 [astro-ph.CO]].

Y.~Rodriguez, J.~P.~B.~Almeida and C.~A.~Valenzuela-Toledo,
``The different varieties of the Suyama-Yamaguchi consistency relation and its violation as a signal of statistical inhomogeneity,''
arXiv:1301.5843 [astro-ph.CO].

D.~H.~Lyth and M.~Karciauskas,
``The statistically anisotropic curvature perturbation generated by $f(\phi)^2 F^2$,''
JCAP {\bf 1305}, 011 (2013)
[arXiv:1302.7304 [astro-ph.CO]].

Tuan Q. Do and W. F. Kao,
``Anisotropic power-law inflation for the Dirac-Born-Infeld theory,''
Phys. Rev. D 84, 123009.\\
Tuan Q. Do, W. F. Kao, and Ing-Chen Lin,
``Anisotropic power-law inflation for a two scalar fields model,'' Phys. Rev. D 83, 123002.

\bibitem{Kim:2013gka}
J.~Kim and E.~Komatsu,
``Limits on anisotropic inflation from the Planck data,''
Phys.\ Rev.\ D {\bf 88}, 101301 (2013)
[arXiv:1310.1605 [astro-ph.CO]].



\bibitem{Soda:2012zm}
J.~Soda,
``Statistical Anisotropy from Anisotropic Inflation,''
Class.\ Quant.\ Grav.\  {\bf 29}, 083001 (2012)
[arXiv:1201.6434 [hep-th]].

\bibitem{Maleknejad:2012fw}
A.~Maleknejad, M.~M.~Sheikh-Jabbari and J.~Soda,
``Gauge Fields and Inflation,''
Phys.\ Rept.\  {\bf 528}, 161 (2013)
[arXiv:1212.2921 [hep-th]].


\bibitem{Emami:2014tpa} 
R.~Emami, H.~Firouzjahi and M.~Zarei,
Phys.\ Rev.\ D {\bf 90}, no. 2, 023504 (2014)
[arXiv:1401.4406 [hep-th]].


\bibitem{Chen:2014vja} 
X.~Chen, R.~Emami, H.~Firouzjahi and Y.~Wang,
JCAP {\bf 1504}, no. 04, 021 (2015)
[arXiv:1408.2096 [astro-ph.CO]].


\bibitem{Ackerman:2007nb} 
  L.~Ackerman, S.~M.~Carroll and M.~B.~Wise,
  Phys.\ Rev.\ D {\bf 75}, 083502 (2007)
  [Erratum-ibid.\ D {\bf 80}, 069901 (2009)]
  [astro-ph/0701357].

\bibitem{Naruko:2014bxa} 
  A.~Naruko, E.~Komatsu and M.~Yamaguchi,
  arXiv:1411.5489 [astro-ph.CO].
  
\bibitem{delta-N}
A. Abolhasani, R. Emami, H. Firouzjahi, unpublished note.   



 \bibitem{Ruth Durrer}
The Cosmic Microwave Background, Ruth Durrer, Cambridge University Press

\end{thebibliography}
\end{document}